\documentclass[10pt,journal,compsoc,cspaper]{IEEEtran}
\usepackage[pdftex]{graphicx}
\usepackage{hyperref}
\usepackage{balance}
\usepackage{cite}
\usepackage{color}
\usepackage{multirow}
\usepackage{amsmath}
\usepackage{array}
\usepackage{hhline}
\usepackage[all=normal,floats,lists,indent,charwidths]{savetrees}
\usepackage{stfloats}

\newcommand{\ignore}[1]{
}

\begin{document}


\title{\huge{Towards Energy-Proportional Computing Using Subsystem-Level Power Management}}

\author{Balaji Subramaniam and Wu-chun Feng 
\IEEEcompsocitemizethanks{\IEEEcompsocthanksitem B. Subramaniam and W. Feng are with 
Department of Computer Science, Virginia Tech. Email: \{balaji, feng\}@cs.vt.edu.}}

\IEEEcompsoctitleabstractindextext{
\begin{abstract}

Massive data centers housing thousands of computing nodes have become 
commonplace in enterprise computing, and the power consumption of such data centers is growing at an
unprecedented rate.  Adding to the problem is the inability of the
servers to exhibit \textit{energy proportionality}, i.e., provide
energy-efficient execution under all levels of utilization, which
diminishes the overall energy efficiency of the data center. It is
imperative that we realize effective strategies to control the
power consumption of the server and improve the energy efficiency of
data centers. With the advent of Intel Sandy Bridge processors, we
have the ability to specify a limit on power consumption during runtime,
which creates opportunities to design new power-management techniques
for enterprise workloads and make the systems that they run on more
\textit{energy-proportional}.

In this paper, we investigate whether it is possible to achieve
\emph{energy proportionality} for enterprise-class server workloads,
namely SPECpower\_ssj2008 and SPECweb2009 benchmarks, by using Intel's Running Average
Power Limit (RAPL) interfaces. First, we analyze the average power consumption 
of the full system as well as the subsystems and describe the energy 
proportionality of these components. We then characterize the 
instantaneous power profile of these benchmarks within different 
subsystems using the \emph{on-chip energy meters} exposed via the RAPL 
interfaces. Finally, we present the effects of power limiting on the energy proportionality, 
performance, power and energy efficiency of enterprise-class server workloads. 
Our observations and results shed light on the efficacy of the RAPL 
interfaces and provide guidance for designing
power-management techniques for enterprise-class workloads.
\end{abstract}

\begin{keywords}
Power Limiting, Energy Proportionality, RAPL, Enterprise Computing, SPEC
\end{keywords}
}

\maketitle

\section{Introduction}

Massive data centers, which house thousands of computing nodes, have become 
increasingly more common. A large fraction of such data centers' total cost of 
ownership (TCO) comes from the cost of building and maintaining infrastructure 
that is capable of powering such large-scale data centers and from the 
recurring energy costs~\cite{provisioning}. Consequently, power and energy 
have emerged as first-order design constraints in data centers. These issues 
are further magnified by the inability of servers to provide 
energy-efficient execution at all levels of utilization (i.e., load-levels). 

\begin{figure}[t]
\centering
\includegraphics[width=1.0\columnwidth]{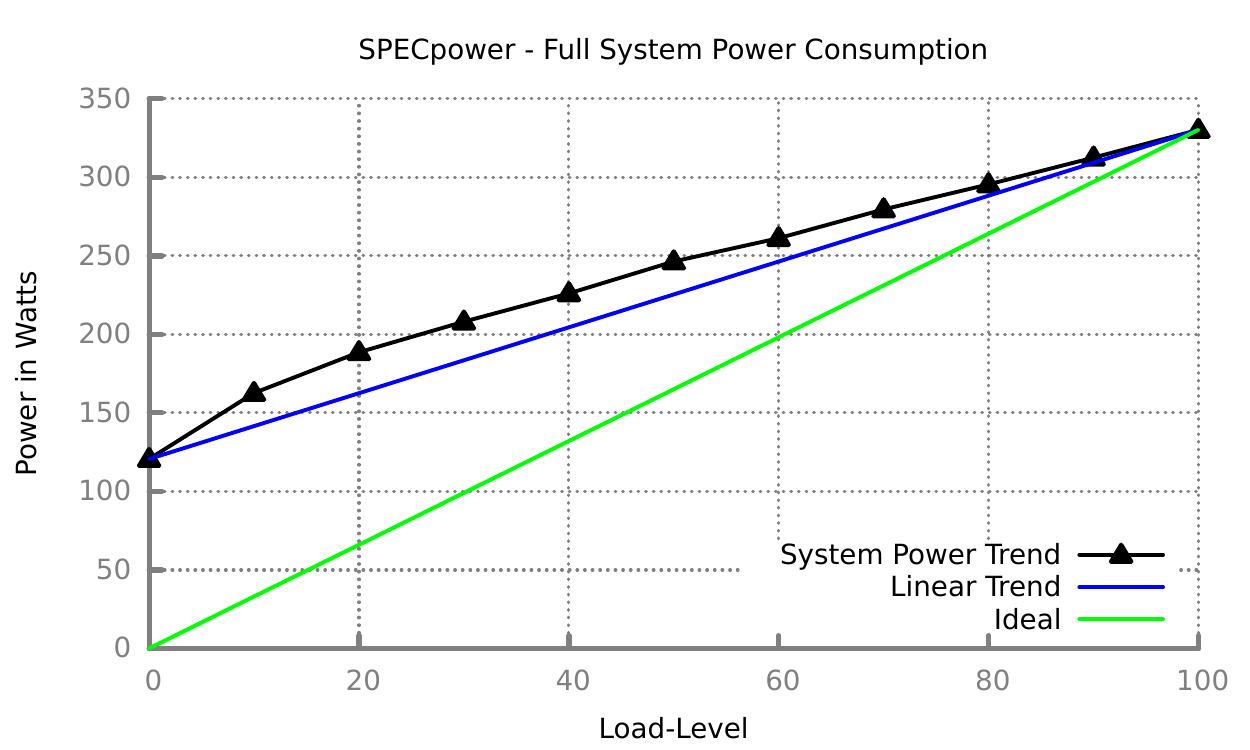}
\caption{Illustration of SPECpower Energy Proportionality}
\label{fig:epgap}
\end{figure}

Figure~\ref{fig:epgap} shows the power consumption of a compute server
running SPECpower under different load-levels and the hypothetical linear and 
ideal (i.e., \emph{energy-proportional}) non-peak power
curves. As evident from the figure, there is room to improve the non-peak power 
efficiency of the server with respect to both the ideal as well as linear power curves. The 
recent recommendation of \emph{energy proportionality} in servers, i.e., to design 
servers that consume power proportional to the utilization, is a move in 
the right direction as it has the potential to double the energy efficiency 
of servers~\cite{eprop}. However, achieving energy-proportional operation is a 
challenging task, particularly given that typical servers consume 35-45\% 
of peak power, even when idling.

Typically, dynamic voltage and frequency scaling (DVFS) has been 
used to achieve better energy efficiency as it can potentially give up to cubic energy 
savings~\cite{beta,adagio,dvfsmyth}. However, as we will show in this paper, the 
subsystem affected by DVFS (i.e., the \emph{core}\footnote{The \emph{core} subsystem 
includes components such as the ALUs, FPUs, L1, and L2 caches~\cite{inteluncore}.}) 
is already the most energy-proportional part of the system. There are other 
subsystems, such as the \emph{uncore},\footnote{The \emph{uncore} subsystem 
includes components such as the memory controller, integrated I/O, and coherence 
engine~\cite{inteluncore}.} that consume constant power, irrespective of 
the system utilization. In order to achieve energy proportionality, we 
need to understand the power consumption of each subsystem at 
different levels of utilization and to leverage mechanisms that enable us to 
control the power consumption of these subsystems. 

With the advent of Intel Sandy Bridge processors, we have better control 
over the power consumption of the system via the Running Average Power 
Limit (RAPL) interfaces~\cite{rapl,intelsdm}. RAPL exposes \emph{on-chip energy meters} 
for the \emph{core} subsystem, processor package, and DRAM and enables the 
tracking of power consumption at a time resolution ($\sim$1 ms) and
system-level granularity that was not possible before. Moreover, it
facilitates deterministic control over the power consumption of 
subsystems through \emph{power limiting} interfaces. These
interfaces allow a user to specify a power bound and a time window
over which the bound should be maintained.  While this
hardware-enforced power limiting is an appealing option, the impact of
power limiting on the performance, power, and energy efficiency of
enterprise-class server workloads is still not well understood and
remains an active area of research.

In this paper, we investigate whether it is possible to achieve
\emph{energy-proportional} operation for enterprise-class server
workloads, namely the SPECpower\_ssj2008 and SPECweb2009 benchmarks (henceforth referred
to as SPECpower and SPECweb respectively) by using the RAPL interfaces.  To this end, this 
paper makes the following contributions: (i) insights into the mechanisms of power management 
for enterprise-class server workloads using the RAPL interfaces via an analysis of the 
SPECpower and SPECweb benchmarks by calibrating its input parameters, (ii) a rigorous 
quantification of the energy proportionality of each subsystem within a server node 
via an analysis of power consumption profiles of the different subsystems when running 
SPECpower and SPECweb at different load-levels, (iii) an analysis and characterization 
of the instantaneous-power profiles at different load-levels of SPECpower and SPECweb to 
understand whether power limiting will enable us to improve the energy efficiency of 
these benchmarks and (iv) empirical results on the impact of RAPL power limiting on 
average power, performance, instantaneous power, and energy efficiency.

Through our contributions, we make the following observations and
conclusions on the power management of the SPECpower and SPECweb benchmarks using RAPL
interfaces:

\begin{itemize}

\item The \emph{core} is the most energy-proportional subsystem
and the \emph{uncore} is the least. 

\item Better power management mechanisms are required to achieve
energy proportionality at the \emph{uncore} subsystem-level.

\item There is ample opportunity for limiting the power consumption of processor 
package and memory subsystems. 

\item Power limiting at the level of the \emph{core} subsystem is the best 
option for improving energy efficiency and achieving energy proportionality. 

\item Though we were not able to achieve energy proportionality at the full system 
level, i.e., entire compute node, we show that
energy-proportional operation or better is possible at the granularity 
of subsystems over which we have control via RAPL power limiting 
(i.e., \emph{core} subsystem, processor package, and DRAM). 

\end{itemize}

The rest of the paper is organized as follows. In
Section~\ref{sec:bkgnd}, we present the details of the SPECpower and SPECweb 
benchmarks and Intel RAPL interfaces.  Section~\ref{sec:avgpower} describes
our analysis and characterization of average power consumption. It presents 
details on the energy proportionality of full system as well as subsytems. 
Section~\ref{sec:instpower} details the instantaneous power profile of 
all subsystems at different load levels in SPECpower and SPECweb and the 
observations from these experiments. Next in Section~\ref{sec:results}, 
we limit the power consumption of SPECpower and SPECweb to study the impact 
of it on the power, performance and energy efficiency of these benchmarks. 
In Section~\ref{sec:related}, we describe the related work, and we
conclude in Section~\ref{sec:conclusion}.


\section{Background}
\label{sec:bkgnd}

In this section, we provide an overview of the SPECpower benchmark and
its design as well as details into its configurable parameters. We
then present 
the control and capabilities exposed by
Intel's RAPL interfaces.

\subsection{Overview of SPECpower Benchmark}
\label{sec:specpower}

SPECpower~\cite{SPECpower} is an industry-standard benchmark that
measures both the power and performance of a server node. The
benchmark mimics a server-side Java transaction processing application. It stresses the
CPU, caches, and memory hierarchy and tests the implementations of the
Java virtual machine (JVM), just-in-time (JIT) compiler, garbage
collection, and threads. The benchmark requires two systems: (1) the
system under test or SUT and (2) the control and collection system
(CCS) with communication between the systems
established via Ethernet.\footnote{SUT and CCS
  can be the same system. Communication is established via Ethernet
  only if the systems are different.} The SUT runs the workload and is
connected to a power meter.  The power meter, in turn, is connected to
the CCS. The CCS collects the performance and power data passed to it
by the SUT and power meter, respectively.

The SPECpower benchmark is designed to produce consistent and
repeatable performance and power measurements. It executes different
type of transactions and the transactions are grouped together in
\textit{batches} for scheduling purposes. Each \textit{load-level} is
achieved by controlling delay between the arrival of batches.  

More
specifically, the SPECpower benchmark is a graduated workload, i.e.,
it runs the workload at different \textit{load-levels} and reports the
power and performance at each
load-level. The benchmark starts with a
calibration phase, which determines the maximum throughput. The
calibrated throughput is set as the throughput target for 100\%
load-level. The throughput target for the rest of the load-levels is
calculated as a percentage of the throughput target for 100\%
load-level. For example, if the throughput target for 100\% load-level
is 100,000, then the target for 70\% load-level is 70,000, 40\% is
40,000 and so on. The throughput is measured in server-side Java
operations per second (ssj\_ops).

The benchmark supports a set of configurable parameters.\footnote{Only 
a subset of these parameters can be changed for compliant runs.} For example,
the maximum target throughput and the batch size can be manually configured. 
We refer the reader to~\cite{SPECpowerrunrules} for further information on configurable parameters.
The flexibility, coupled with the consistency and repeatability of SPECpower,
allows us to evaluate the applicability of newer power-management interfaces,
such as RAPL, to enterprise-class server workloads.

\subsection{Overview of SPECweb Benchmark}
\label{sec:specweb}

SPECweb is an industry standard benchmark for measuring front-end web server performance. 
It allows the user to measure performance based on the request handling capability and response 
time maintained by a server node. The benchmark consists of four different components:

\begin{enumerate}

\item Client: It runs the application program which sends HTTP 
requests to the server and receives the corresponding response from the 
server. 

\item Web Server: It handles the requests issued by the client. This is also the 
system under test (SUT) for this benchmark.

\item Back-End Simulator (BeSim): It emulates a back-end application server. The web 
server communicates with Besim to retrieve specific information required to complete a
request from one of the clients.

\item Prime Client: It initiates and controls the clients and also 
initializes the web server and Besim. It collects performance results for the benchmark.
\end{enumerate} 

The main performance and power metric for the benchmark is simultaneous user 
sessions (SUS) and SUS/watt respectively. In addition to SUS, the SPECweb 
benchmarks adds two different response time performance metrics, namely 
\emph{TIME\_GOOD} and \emph{TIME\_TOLERABLE}. By default, 95\% and 99\% of the requests 
should have response time less that \emph{TIME\_GOOD} and \emph{TIME\_TOLERABLE} 
respectively. Similar to SPECpower, we can control the benchmark 
parameters to execute the benchmark at different \emph{load-levels} i.e., different 
SUS. This benchmark also allows us tweak a set of input parameters. 
We refer the reader to~\cite{specwebguide} for a full list of configurable parameters. 


\subsection{Intel's Running Average Power Limit (RAPL) Interfaces}
\label{sec:rapl}

RAPL was introduced in Intel Sandy Bridge processors. The RAPL interfaces
provide mechanisms to enforce power consumption limits on a specific subsystem. 
The only official documentation available for these interfaces is section 14.7 of the Intel software 
developer's manual~\cite{intelsdm}. Our experiments deal only with the Sandy 
Bridge server platforms.

The RAPL interfaces can be programmed using the model-specific registers 
(MSRs). MSRs are used for performance monitoring and controlling hardware 
functions. These registers can be accessed using two instructions: (1) 
\textit{rdmsr}, short for ``read model-specific registers'' and 
(2) \textit{wrmsr}, short for ``write model-specific registers.'' 
The \textit{msr} kernel module can be used for accessing 
MSRs from \textit{user space} in Linux environments. When loaded, the 
\textit{msr} module exposes a file interface at \textit{/dev/cpu/x/msr}. 
This file interface can be used to read from or write to any MSR on that 
CPU.

According to the Intel documentation, RAPL interfaces operate at the
granularity of a processor socket. The server platforms provide
control over three domains (i.e., subsystems):\footnote{Note: We use
  RAPL domain and subsystem interchangeably in rest of the paper.}
(1) package (PKG), (2) power plane 0 (PP0), and (3) DRAM. PKG, PP0 and DRAM 
represents the processor package (or socket), the
\emph{core} subsystem, and memory DIMMs associated with that socket,
respectively.  The \textit{MSR\_RAPL\_POWER\_UNIT} register contains
the units for specifying time, power, and energy, and the values are
architecture-specific. For example, our testbed requires and reports
time, power, and energy at increments of 976 microseconds, 0.125 watts,
and 15.3 microjoules, respectively.  Each domain consists of its own set
of RAPL MSR interfaces. On a server platform, RAPL exposes four
capabilities:

\begin{enumerate}
\item Power limiting -- Interface to enforce limits on power consumption.
\item Energy metering -- Interface reporting actual energy usage information.
\item Performance status -- Interface reporting performance impact due to power limit. 
\item Power information -- Interface which provides value range for control 
attributes associated with power limiting. 
\end{enumerate}

\subsubsection{Power Limiting}

RAPL maintains an average power limit over a sliding window 
instead of enforcing strict limits on the instantaneous power. The advantage 
of having an average power limit is that if the average 
performance requirement is within the specified power limits the workload 
will not incur any performance degradation even if the performance 
requirement well exceeds the power limit over short bursts of time. The 
user has to provide a power bound and a time window in which the limit has 
to be maintained. Each RAPL domain exposes a MSR which is used for programming 
these values. The PKG domain provides two power limits and associated 
time window for finer control over the workload performance whereas other 
domains provide only one power limit. The interface provides a \textit{clamping} 
ability, which when enabled, allows the processor to go below an OS-requested P-state. 

\subsubsection{Energy Metering}

Each domain exposes a MSR interface that reports the energy consumed
by that domain. On a server platform, (1)
\textit{energy(PKG) = energy consumed by the processor package}, (2)
\textit{energy(PP0) = energy consumed by the core subsystem}, and (3)
\textit{energy(uncore subsystem) = energy(PKG) $-$ energy(PP0)}.

\subsubsection{Performance Status}

This MSR interface reports the total time for which each domain was 
throttled (i.e., functioning below the OS-requested P-state) due to the 
enforced power limit. This information will be useful in understanding 
the effects of power limiting on a particular workload. 

\subsubsection{Power Information}
\label{sec:powerinfo}

The PKG and DRAM domains expose a MSR interface that provides
information on the ranges of values that can be specified for a
particular RAPL domain for limiting its power consumption. This
includes maximum time window, maximum power, and minimum power. The
range of per-socket values on our experimental platform is given in
Table~\ref{tab:raplrange}.

\begin{table}
\centering

\caption{Per-Socket Parameter Range (MTW = Maximum Time Window, MaxP = Maximum 
Power, MinP = Minimum Power). Multiply by 2 for Full Two-Socket System.}
\label{tab:raplrange}

\begin{tabular}{|c|c|c|c|} 
\hline
Domain/Range & MTW & MaxP & MinP \\ \hline \hline
Package & 45.89 ms & 180 watts & 51 watts \\ \hline
DRAM & 39.06 ms & 75 watts& 15 watts\\ \hline
\end{tabular}

\end{table}

\section{Experimental Setup}
\label{sec:setup}

The SUT for our experiments is an Intel Xeon E5-2665 processor 
(Intel Romley-EP). The node has two such processors for a 
total of 16 cores and 32 cores when hyperthreading is ON. 
It has 256 GB of memory and runs a Linux kernel version 3.2.0. We used a 
Yokogawa WT210 power meter for full system power measurements. 

\subsection{Setup for SPECpower}
The CCS has an Intel Xeon E5405 processor with 
dual quad cores and 8 GB of RAM. The CCS runs a Linux kernel version 2.6.32. 
The CCS and SUT were connected through a gigabit Ethernet network. 
We used all the cores in SUT for our experiments. Eight JVMs with four 
threads for each JVM were used as the configuration for
SPECpower. The four threads in each JVM were pinned to two
adjacent physical cores on the SUT using \emph{numactl}. To further enhance the 
performance of the SUT, we enabled large page memory (HugeTLB) support 
and set aside 32 GB for huge page allocation. Note that HugeTLB support 
is enabled only for SPECpower. In order to
provide consistent performance results throughout our experiments, we
configured the \textit{input.load\_level.target\_max\_throughput}
parameter to achieve the same performance for each run. It was set to
140,000 ssj\_ops for each JVM for a total of 1,120,000 ssj\_ops for
the entire run. In all our experiments, 100\% load-level corresponds
to 1,120,000 ssj\_ops. This value was determined by averaging 10
calibration runs. We changed the runtime for each load-level to 120 seconds using the
\emph{input.load\_level.length\_seconds} parameter and the pre- and
post-measurement interval to 15 seconds in order to reduce the total
runtime of the benchmark. We use 1000 as our batch size as there is minimal 
to no effect on power due to batch sizes (See Appendix~\ref{sec:app1}). On an average, the SUT consumes 120 watts
at idle and 330 watts at 100\% load-level of SPECpower. We would like to 
stress that the system consumes 36.51\% of peak power\footnote{Power 
consumed at 100\% load-level.} even when idling.

\subsection{Setup for SPECweb}

We used 26 clients, 1 prime client and 2 Besim for our experiments. The prime 
client is an Intel Xeon E5405 processor with two quad cores and 
8 GB of RAM. The Besims had two dual core AMD Opteron 2218 processors with 4 
GB of RAM. In this paper, we benchmark only the SPECweb\_PHP\_Ecommerce 
workload. We used a Apache installation 
with php module as our web serving application. We setup a bonded Ethernet link with the 
available ports on the SUT to enable data transmission upto 2 Gbps. Note that the 
bonded Ethernet link is only setup for SPECweb. In our experiments for SPECweb, 
100\% load-level corresponds to 13000 SUS. This value was determined using 
empirical analysis (see Appendix~\ref{sec:app2}). In addition to the sessions, all our experiments also 
maintain the response time criteria. In our case, 95\% (\emph{TIME\_GOOD} parameter) 
and 99\% (\emph{TIME\_TOLERABLE} parameter) of the requests need to have 
response times less than 3 and 5 seconds, respectively. These response time constraints are default 
values and used in the compliant runs. The load-level is changed by manually 
modifying the SIMULTANEOUS\_SESSIONS parameter in the input configuration. 
We modified the RUN\_SECONDS input parameter 
to 420 seconds to reduce the runtime of the benchmark. Since we focus only on the processor package 
and memory power management, we load all the data associated with the Ecommerce 
workload into RAMFS to keep the data set in memory and minimize the involvement 
of disks. On an average, the SUT consumes 120 watts when idling and 219 watts 
at 100\% load-level of SPECweb. In case of SPECweb, the system consumes 54.88\% 
of peak power when idling. 

\section{An Analysis of Average Power Consumption}
\label{sec:avgpower}

In this section, we characterize the power consumption of the SPECpower and SPECweb benchmarks 
and analyze energy proportionality from the perspective of the entire system 
as well as each RAPL domain. Through our experiments, we will show that 
the most and least energy-proportional subsystems are the \emph{core} (PP0) and the
\emph{uncore} (Package-PP0), respectively. 


\subsection{Power Consumption Analysis}

As discussed earlier, we are interested in analyzing the energy
proportionality of the system. The deviation of the power curve of the
system from the ideal power curve is of particular interest to us. To
illustrate with an example, we would like the area between the system and 
the ideal power curve to be as small as possible in Figure~\ref{fig:epgap}.  
Henceforth, this area will be referred to as \textit{energy proportionality gap (EPG)}.
We are also interested in the linearity of the system power curve which is the 
area between the system power trend and linear curve. Henceforth, this area will 
be referred to as \textit{linear deviation gap (LDG)}.

\subsubsection{Properties of Energy-Proportional Systems}
\label{sec:epprop}

Barroso et. al~\cite{eprop} advocated the design of energy-proportional 
systems by addressing power characteristics of the server and the behavior of 
enterprise-class server workloads. They proposed two properties of 
energy-proportional systems -- low idle power and wide dynamic power range. 
These two properties are particularly illustrated by the ideal curve in 
Figure~\ref{fig:epgap}. The ideal curve consumes zero power when idling 
(i.e., at 0\% load-level) and has a wide dynamic power range. In this paper, 
we will quantify the idle power and the dynamic power range as percentage 
of peak power (i.e., the power consumed at 100\% load-level). 

\subsubsection{Energy Proportionality Metric}

We quantify the EPG using two different metrics: (1) the EP metric~\cite{epmetric} 
and (2) the PG metric~\cite{knightshift}. Each of these metrics serve 
different purposes and quantifies the energy proportionality of the system along 
different granularities. The EP metric is calculated as shown in Equation~\ref{eq:epmetric} 
where $Area_{System}$ and $Area_{Ideal}$ represent the area under the system and ideal 
power curve respectively. A value of~1 for the metric represents an 
ideal energy-proportional system.  A value of~0
represents a system that consumes a constant amount of power
irrespective of the load-level. A value greater than~1 represents a system 
which is better than energy-proportional.\footnote{Originally, the EP 
metric proposed in~\cite{epmetric} varied only between 0 and 1 (i.e., it did 
not account for better than energy-proportional systems). However, in 
this paper we extend EP metric to account for better than energy-proportional 
system (i.e., 0 $<$ EP metric $<$ 2).} The EP metric gives 
a perspective of the energy proportionality of the system at the full system 
level. 
\begin{equation}
\label{eq:epmetric}
EP = 1-\frac{Area_{System} - Area_{Ideal}}{Area_{Ideal}}
\end{equation}

The PG metric is calculated as shown in Equation~\ref{eq:pgmetric} where $X\%$ represents $X\%$ 
load-level. As observed, the PG metric defines the EPG at individual load-levels. For an ideal 
energy-proportional server, the PG for all utilization should be 0. 

\begin{equation}
\label{eq:pgmetric}
PG_{X\%} = \frac{Power_{System@X\%} - Power_{Ideal@X\%}}{Power_{System@100\%}}
\end{equation}

The LDG is quantified using LD metric~\cite{knightshift}. The LD metric 
is calculated using Equation~\ref{eq:ldmetric}. For an linear energy-proportional system, 
the LD metric will be 0. LD metric $>$ 1 and $<$ 1 indicate superlinear 
and sublinear energy proportional systems. 

\begin{equation}
\label{eq:ldmetric}
LD = \frac{Area_{System}}{Area_{Ideal}} - 1
\end{equation}

We will use the properties described in Section~\ref{sec:epprop} along with 
the EP metric, PG metric and LD metric to quantify the energy proportionality. 
We will also look at the EPG and LDG both at full system- and subsystem-levels in rest of the sections. 

\subsubsection{Methodology}
We used the energy meters exposed in each RAPL domain to determine the
power dissipated in each domain. In all our results, we report the
average power\footnote{Average power is calculated as (initial energy
  reading - final energy reading)/time.}  over ten runs for the
domain-level power consumption. For full-system power measurement, we
have followed the power measurement methodology specified and
developed by the SPEC organization for the SPECpower and SPECweb 
benchmarks~\cite{SPECpowermethod}.

\subsubsection{Analysis of System- and Subsystem-Level Energy Proportionality}

In this section, we present the details on the power consumption of SPECpower and SPECweb 
at a subsystem-level. We were able to profile the benchmark at a 
granularity that has not been possible until the advent of Intel Sandy Bridge 
by using the on-chip energy meters exposed by the RAPL interfaces. 
Our results provide insights into the energy proportionality of a
system as a whole as well as at the RAPL domain-level.

\begin{figure}[h]
\centering
\includegraphics[width=1.0\columnwidth]{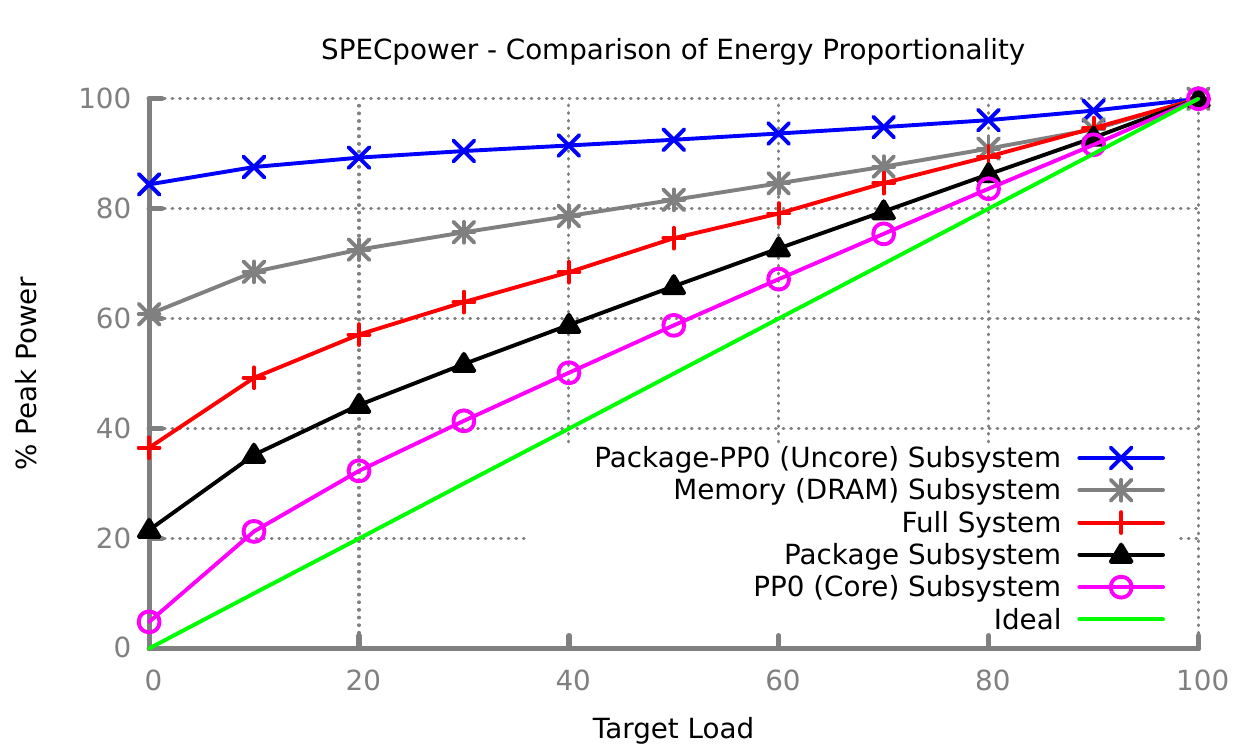}
\caption{Analysis of SPECpower Energy Proportionality}
\label{fig:epcomparison1}
\end{figure}

\begin{figure}[t]
\centering
\includegraphics[width=1.0\columnwidth]{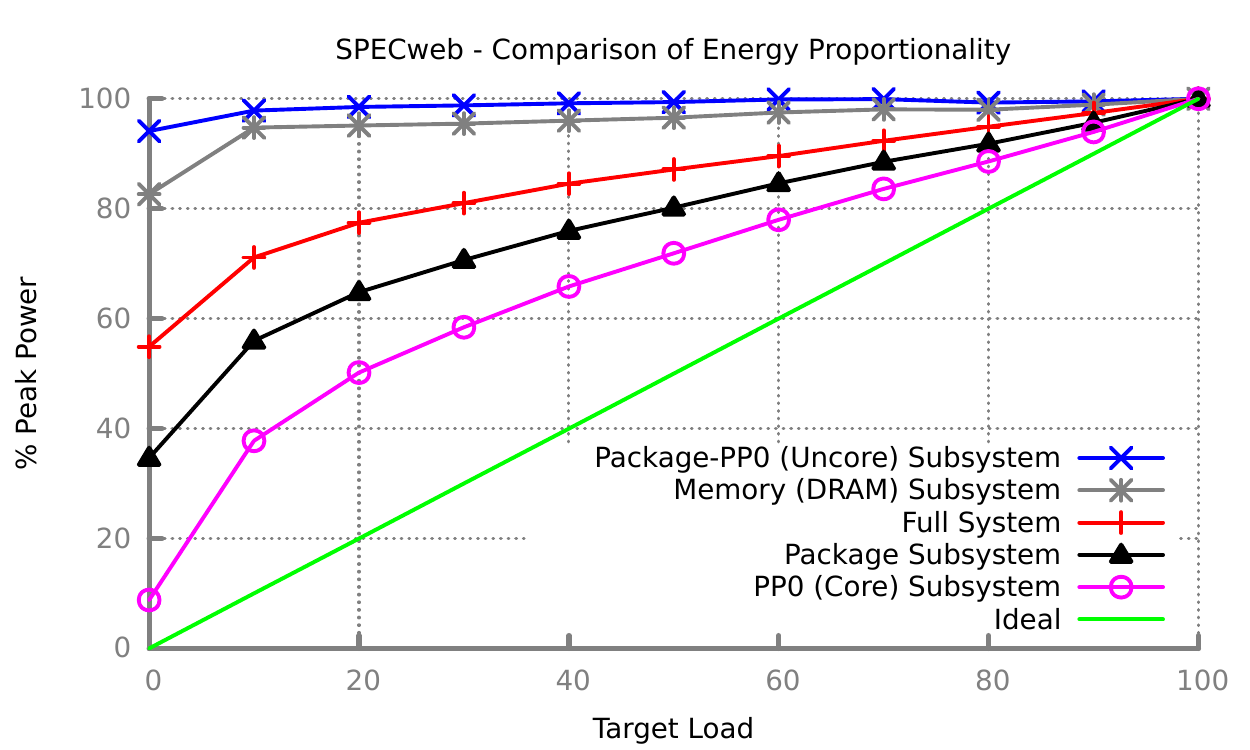}
\caption{Analysis of SPECweb Energy Proportionality}
\label{fig:epcomparison2}
\end{figure}

Figures~\ref{fig:epcomparison1} and~\ref{fig:epcomparison2} describes the energy proportionality of the full 
system and different subsystems. The Y-axis represents 
the percentage of peak power consumed by the system or 
subsystem and X-axis represents the load-level. As a 
result, the ideal curve (green line) consumes 40\% of peak power at 40\% load-level, 
60\% of peak power at 60\% load-level and so on. Figures~\ref{fig:epcomparison1} and~\ref{fig:epcomparison2} are also a 
compact comparison of the energy proportionality of different components of 
the system. As mentioned earlier, we will quantify the energy proportionality using the EP, PG and LD 
metrics and the desired properties of an energy-proportional system. 

\paragraph{Full System Energy Proportionality}

The energy proportionality of full system is represented by the red line 
in Figures~\ref{fig:epcomparison1} and~\ref{fig:epcomparison2}. The EP metric for full system is 0.54 and 
0.29 for SPECpower and SPECweb respectively. Full system idles at 36.51\% and 54.88\% of peak power for 
SPECweb and SPECpower. Therefore, it is impossible to achieve energy-proportional 
operation for load-levels less than 36\% in case of SPECpower and 54\% in case of SPECweb. The dynamic power range 
is 63.48\% for SPECpower and 45.11\% for SPECweb.\footnote{Dynamic power 
range is calculated as power consumed at 100\% load-level - 0\% load-level.}   

\paragraph{Package (PKG) Energy Proportionality} 
The EP metric value for the package subsystem is 0.70 and 0.44 for SPECpower and SPECweb respectively. 
It is also worth noting that the power profile of package and full system 
follow a similar trend for both the benchmarks, indicating a strong correlation 
between them.\footnote{The Pearson correlation is greater than 0.99.} The package subsystem idles 
at 21.55\% and 34.47\% for SPECpower and SPECweb respectively. The dynamic power 
range for SPECpower is 78.44\% and SPECweb is 65.52\%. In general, due to its better 
EP metric, lower idle power and high dynamic power range package subsystem is \emph{more} 
energy-proportional than the full system. 

\paragraph{\emph{Core} (PP0) Energy Proportionality}

The purple line in Figures~\ref{fig:epcomparison1} and~\ref{fig:epcomparison2} describes the energy proportionality of 
the PP0 domain. We observe that this subsystem has near energy-proportional power 
profile for SPECpower benchmark. However, it is relatively less energy-proportional 
in case of the SPECweb benchmark. It has a EP metric value of 0.85 in case of SPECpower 
and 0.63 in case of SPECweb. This subsystem idles at 5.74 watts (4.83\% 
of peak power) and has a dynamic power range of 95.16\% of peak power for SPECpower. The 
idle power and dynamic power range are 8.80 and 91.19 percent of peak power for the SPECweb 
benchmark. The low idle power coupled with the high dynamic power range makes 
this subsystem suitable to be operated at different power-performance trade-offs. 

\paragraph{\emph{Uncore} (Package-PP0) Energy Proportionality}

The \textit{uncore} subsystem's power consumption remains almost constant irrespective 
of the load-level with an EP metric value of 0.14 for SPECpower and 0.02
for SPECweb. The \textit{uncore} subsystem has the greatest EPG, and as 
a result, exhibits the worst power consumption trend among the 
full system and RAPL domains from the perspective of energy-proportional 
power scaling. It idles at 84.41\% and 94.13\% of peak power for SPECpower 
and SPECweb, respectively. It has the least dynamic power range among all 
systems and subsystems at 15.58\% for SPECpower and 5.86\% for SPECweb.

\paragraph{Memory (DRAM) Energy Proportionality}

The memory subsystem has EP metric value of 0.36 for SPECpower and 0.07 for SPECweb. 
In case of SPECweb, the memory power trend closely follows the uncore power trend 
which makes it less energy-proportional. This worse memory energy proportionality of 
the SPECweb benchmark can be attributed to the usage of RAMFS to house 
the data required by the web server. This subsystem idles at 
60.80\% and 82.62\% for SPECpower and SPECweb. 

To summarize, Table~\ref{tab:epropanalysis} describes our results on the energy proportionality 
analysis of full system and subsystems.  

\begin{table*}[htb]

\centering
\caption{Summary of Full System- and Subsystem-Level Energy Proportionality Analysis. Note: Idle Power and Dynamic Power Range are Represented as 
Percentage of Peak Power.}
 \label{tab:epropanalysis}

\begin{tabular}{|c|c|c|c|c|} 
\hline
\textbf{Subsystem}	&	\textbf{Benchmark} & \textbf{EP Metric}	& \textbf{Idle Power} 	& \textbf{Dynamic Power Range} \\ \hline
\multirow{2}{*}{Full System}	&	SPECpower & 0.54	& 36.51	& 	63.48	\\ \hhline{~----}
								&	SPECweb & 0.29	& 54.88	&	45.11	\\ \hline
\multirow{2}{*}{Package (PKG)} 	&	SPECpower &	0.70	& 21.55	& 78.44	\\ \hhline{~----}
								&	SPECweb & 0.44	& 34.47	& 	65.52	\\ \hline
\multirow{2}{*}{\emph{Core} (PP0} 	&	SPECpower &	0.85	&	4.83	& 	95.16	\\ \hhline{~----} 
								&	SPECweb & 0.63	& 8.80	& 	91.19	\\ \hline
\multirow{2}{*}{\emph{Uncore} (Package-PP0)} 	&	SPECpower &	0.14	& 84.41	& 15.58	\\ \hhline{~----}
												&	SPECweb & 0.02	& 94.13	& 	5.86	\\ \hline
\multirow{2}{*}{Memory (DRAM)} 	&	SPECpower &	0.36	& 60.80	& 39.19	\\ \hhline{~----}
								&	SPECweb & 0.07	& 82.62	&	17.37	\\ \hline
\end{tabular}
\end{table*}

\subsubsection{Analysis of Load-Level Energy Proportionality}

\begin{figure*}[t]
\centering
\includegraphics[width=1.0\columnwidth]{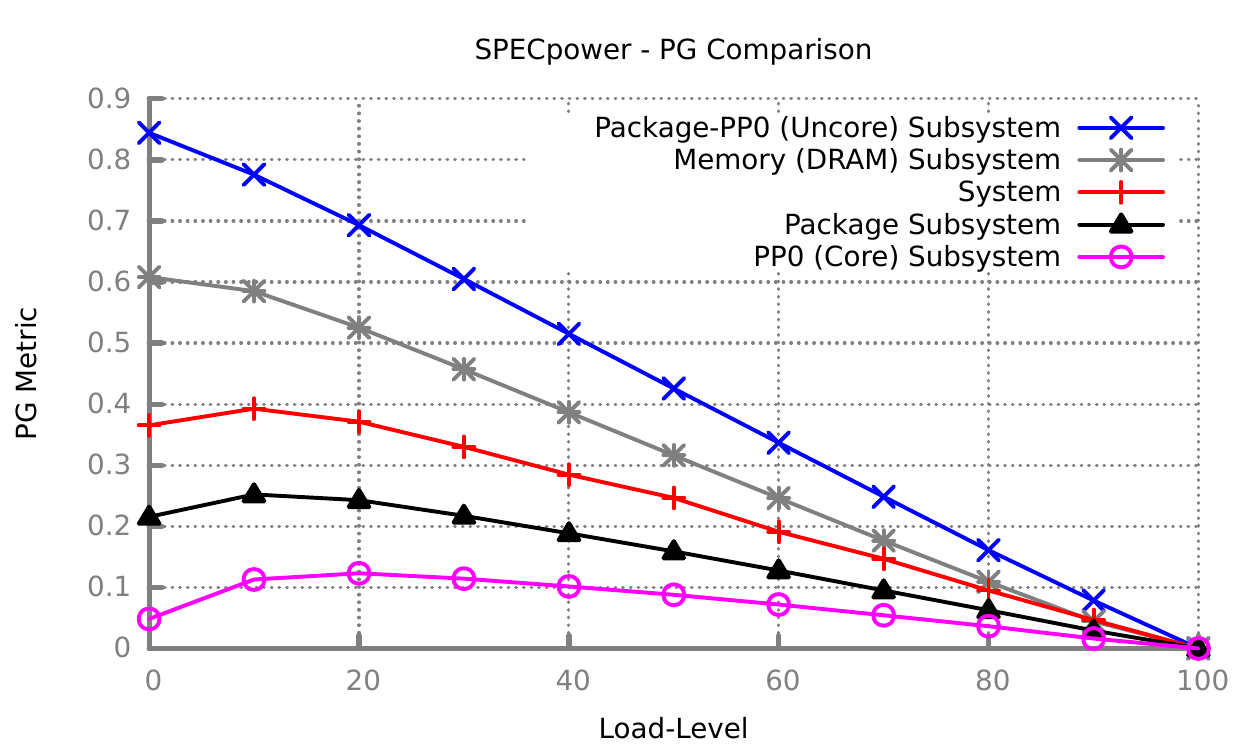}
\includegraphics[width=1.0\columnwidth]{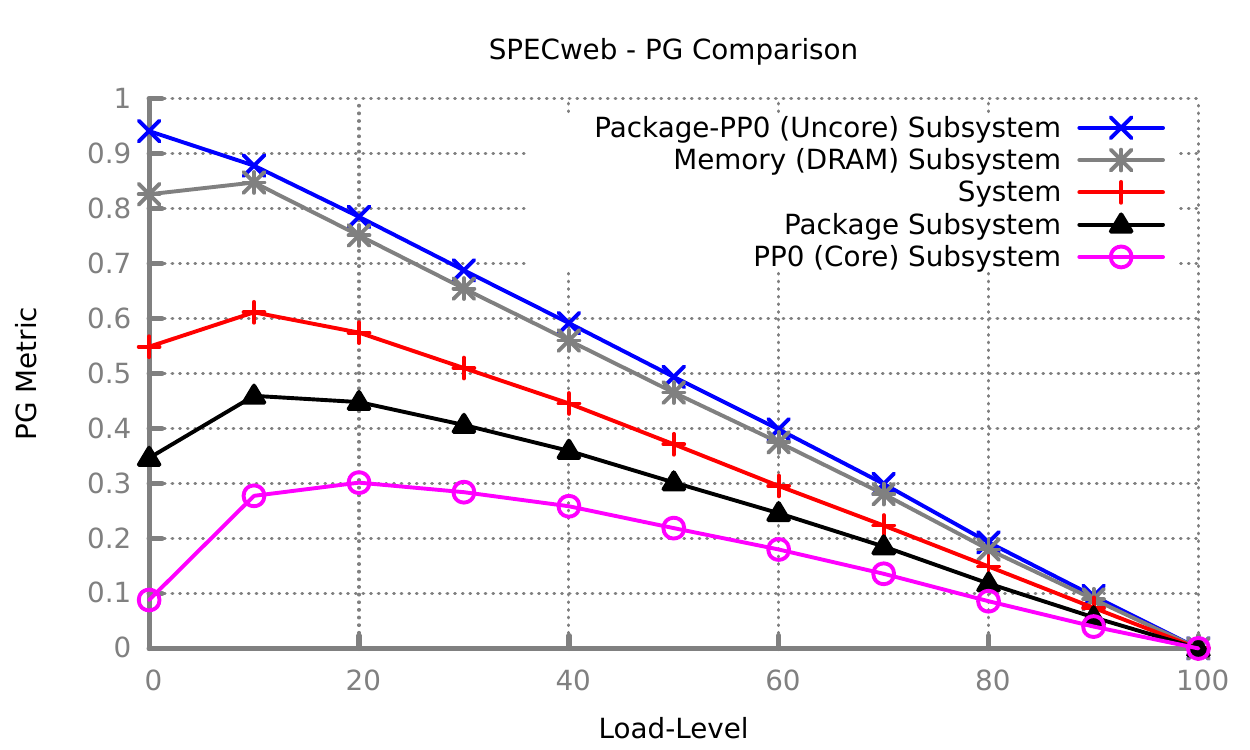}
\caption{Analysis of Load-Level Energy Proportionality}
\label{fig:pgcomparison}
\end{figure*}

The PG metric allows us to look at the energy proportionality of a server at 
each load-level. Figure~\ref{fig:pgcomparison} shows the PG metric at each load-level for SPECpower and 
SPECweb benchmarks. Similar to EP metric, the \emph{uncore} and \emph{core} subsystem 
have the worst and best PG metric for all load-levels for both the benchmarks. The \emph{uncore} 
subsystem's PG increases linearly from 100 to 0 percent load-level which again shows that 
the subsystem's power remains a constant irrespective of the load-level. In case of 
the PP0 subsystem, there is an increase in PG metric when load-level increases from 0 to 10 
percent. This trend shows that the energy proportionality gap at 0\% load is better than low 
utilization levels for both the benchmarks. The proportionality 
gap becomes better than 0\% load-level only at 70\% and 80\% load-level for SPECpower 
and SPECweb benchmarks, respectively, for the PP0 subsystem. Such trends 
can be seen for Package subsystem and full system as well. 

\emph{In summary, core is the most energy-proportional and the uncore 
is the least energy-proportional subsystem.}

\subsubsection{Analysis of Linear Deviation}

Table~\ref{tab:ldanalysis} shows the LD metric for both benchmarks at each RAPL 
domain and full system. The LD metric for all subsystems is always positive as
none of them have a sub-linear energy proportionality trend. This observation also 
provides evidence that there is opportunity to improve the energy proportionality 
by improving (i.e., decreasing) LD metric. 

\begin{table}[h]

\centering
\caption{Summary of Full System- and Subsystem-Level Linear Deviation Analysis.}
\label{tab:ldanalysis}

\begin{tabular}{|c|c|c|} 
\hline
\textbf{Subsystem}	&	\textbf{Benchmark} & \textbf{LD Metric} \\ \hline
\multirow{2}{*}{Full System}	&	SPECpower & 0.067	\\ \hhline{~--}
								&	SPECweb & 0.101	\\ \hline
\multirow{2}{*}{Package (PKG)} 	&	SPECpower &	0.066 \\ \hhline{~--}
								&	SPECweb & 0.151	\\ \hline
\multirow{2}{*}{\emph{Core} (PP0} 	&	SPECpower & 0.095	\\ \hhline{~--} 
								&	SPECweb & 0.254	\\ \hline
\multirow{2}{*}{\emph{Uncore} (Package-PP0)} 	&	SPECpower &	0.004	\\ \hhline{~--}
												&	SPECweb & 0.019	\\ \hline
\multirow{2}{*}{Memory (DRAM)} 	&	SPECpower &	0.013	\\ \hhline{~--}
								&	SPECweb & 0.053	\\ \hline
\end{tabular}
\end{table}

\section{An Analysis of Instantaneous Power Consumption}
\label{sec:instpower}

Here we present our results for the instantaneous power profile
analysis of the SPECpower and SPECweb benchmarks. Our main goal is to 
visualize the opportunities for power limiting. We collected
instantaneous power profile for five load-levels. 

\subsection{Methodology}

Our results are shown as cumulative distribution functions (CDFs). The 
CDFs present the percentage of time spent at or below a given percentage 
of the maximum power limit possible. We refer the reader to Table~\ref{tab:raplrange} 
for the maximum power limit possible for each subsystem. We collect the 
instantaneous power profile of the package and memory subsystems at 50 ms 
resolution. The results are normalized to their respective maximum power 
limit possible. 

\subsection{Instantaneous Power Analysis for Package (PKG) Subsystem}

\begin{figure}[h]
\centering
\includegraphics[width=1.0\columnwidth]{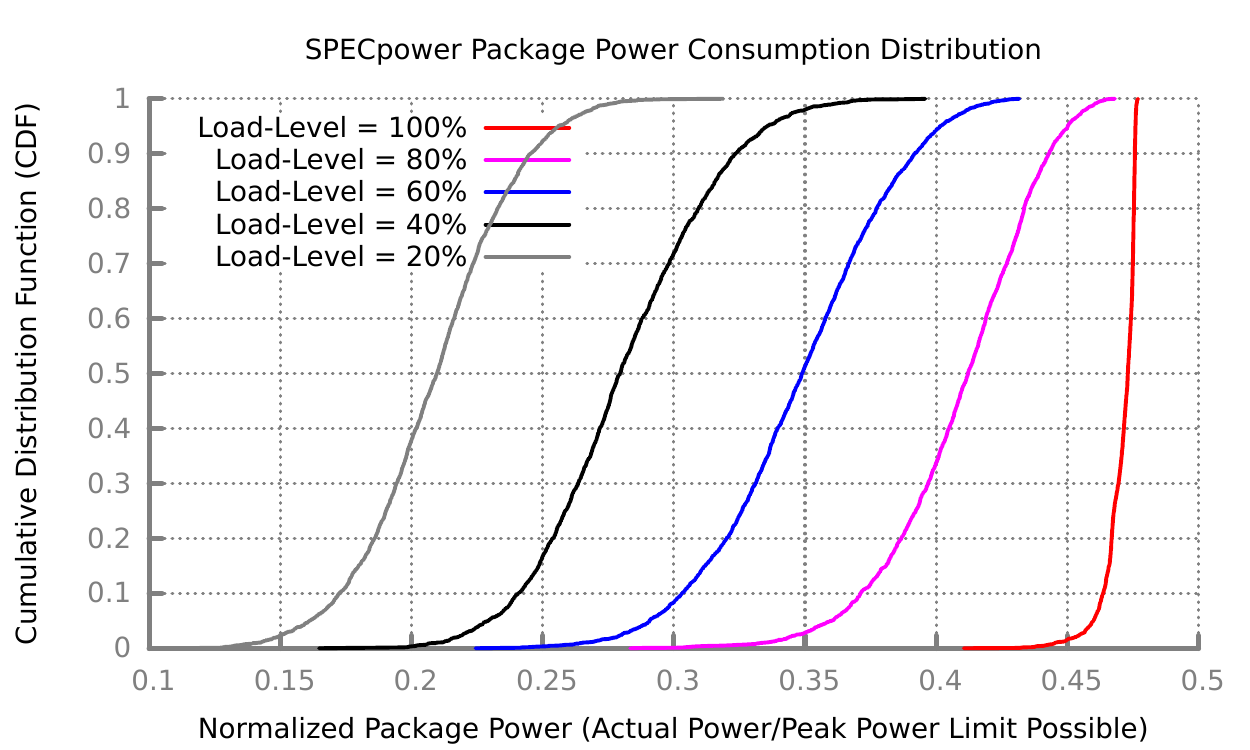}
\caption{Analysis of SPECpower Instantaneous Power Consumption For Package (PKG) Subsystem}
\label{fig:instpkg1}
\end{figure}

\begin{figure}[h]
\centering
\includegraphics[width=1.0\columnwidth]{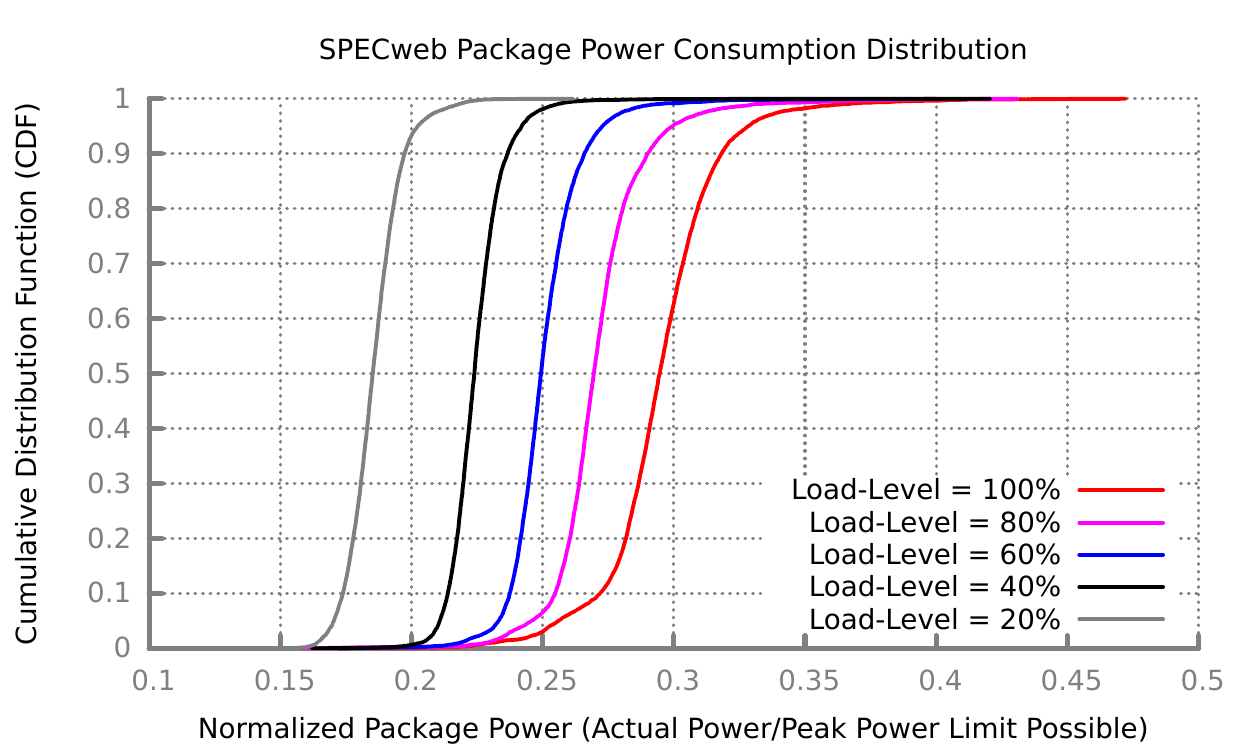}
\caption{Analysis of SPECweb Instantaneous Power Consumption For Package (PKG) Subsystem}
\label{fig:instpkg2}
\end{figure}

Figures~\ref{fig:instpkg1} and~\ref{fig:instpkg2} show the instantaneous power consumption for package 
subsystem for five different load-levels of SPECpower and SPECweb. 
We observe that the maximum power consumed by 100\% load-level of SPECpower 
as indicated by its CDF is lower than 0.5 normalized power. 
This indicates that the maximum power consumed 
while executing SPECpower is less than 50\% of maximum power limit possible. 
This upper limit for package power consumption for the SPECweb benchmark is 
also less than 50\% of the maximum power limit possible.  
The lowest point in the CDF of each workload corresponds to the minimum 
power consumed. At no point during the execution of that load-level, the 
subsystem consumes lesser power. For example, 100\% and 60\% load-levels of 
SPECpower do not consume less than 40\% and 20\% of normalized power respectively. The shape of 
the curves indicate that each load-level spends most of the time consuming a 
narrow range of power. For instance in case of SPECpower, 80\% load-level spends most of the time 
consuming power between 34\% and 46\% of maximum power limit possible. For both the 
benchmarks, we will benefit by removing the relatively few intervals (indicated 
by the flat lines at 100\%) where the workload has a power spike. Power limiting can help 
in such cases to remove these few intervals. 
We also observe that the power range decreases with increase in 
load-level. In case of SPECpower, 100\% load-level has a power range from 0.40 to 0.50 of the normalized 
power whereas 20\% load-level has a power range from 0.10 to 0.32 of the normalized 
power.

\subsection{Instantaneous Power Analysis for Memory (DRAM) Subsystem}

\begin{figure}[h]
\centering
\includegraphics[width=1.0\columnwidth]{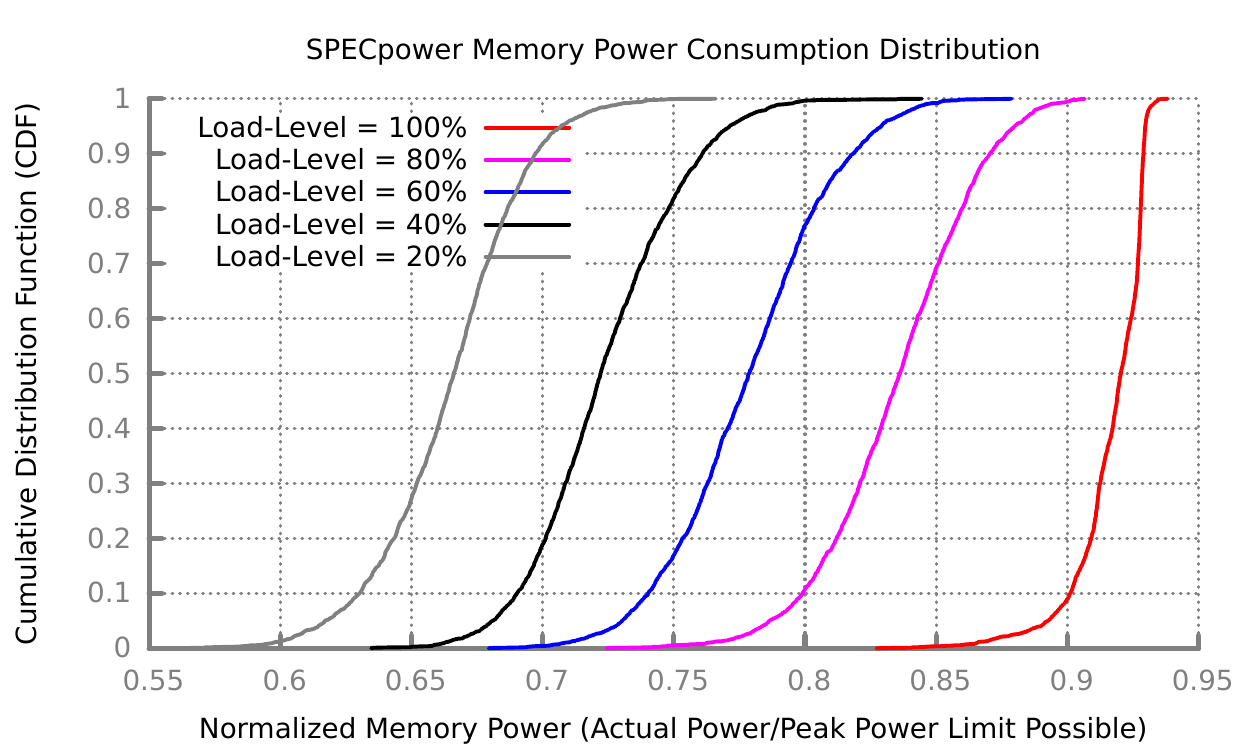}
\caption{Analysis of SPECpower Instantaneous Power Consumption For Memory (DRAM) Subsystem}
\label{fig:instmem1}
\end{figure}

\begin{figure}[h]
\centering
\includegraphics[width=1.0\columnwidth]{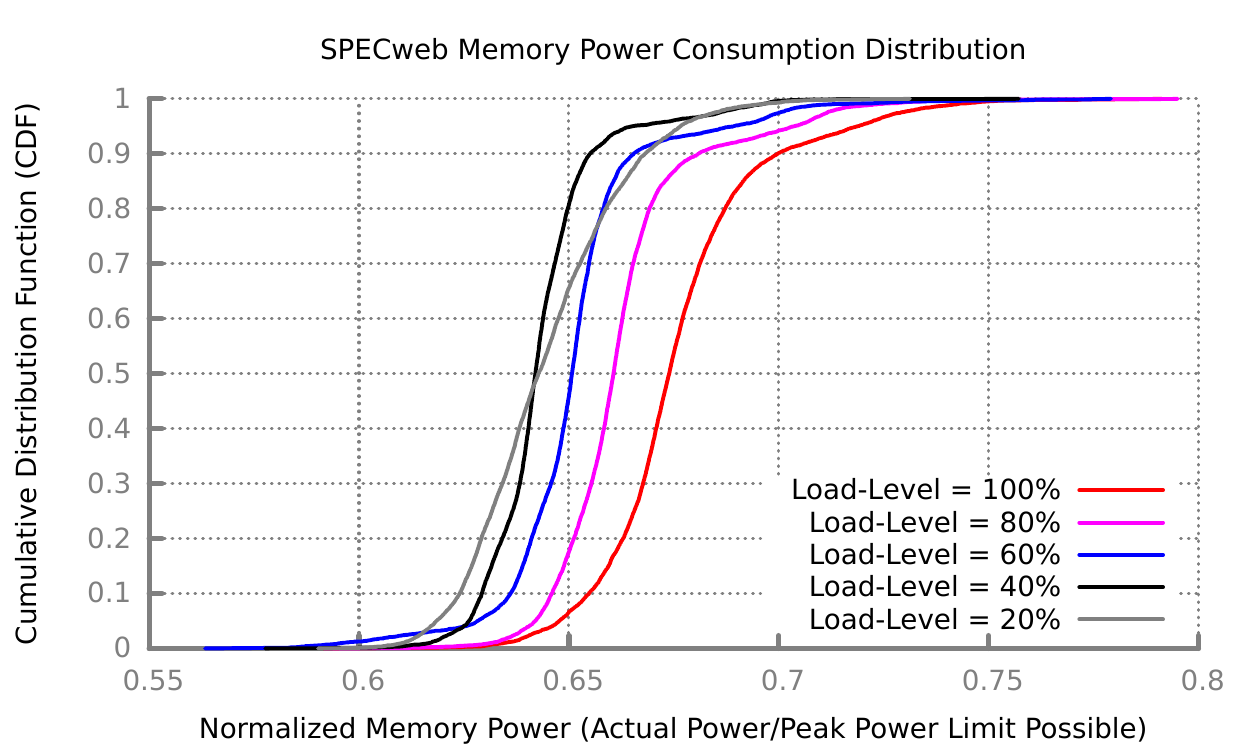}
\caption{Analysis of SPECweb Instantaneous Power Consumption For Memory (DRAM) Subsystem}
\label{fig:instmem2}
\end{figure}

Figures~\ref{fig:instmem1} and~\ref{fig:instmem2} describe the instantaneous power consumption for 
memory subsystem for five different load-levels of SPECpower and SPECweb respectively. We observe 
CDF curves similar to package subsystem for both the benchmarks. Minimum normalized power consumed 
at each load-level is higher than the corresponding observation for package 
subsystem. This is an expected behavior as memory subsystem idles at a 
higher percentage of peak power than the package subsystem 
(see Table~\ref{tab:epropanalysis}). Similar to package subsystem, each 
load-level spends most of the time consuming a narrow range of power. The 100\% 
load-level for SPECpower consumes 87 to 93 percent of peak power limit possible 
leaving lesser opportunity for memory power management than other load-levels. The 
memory power consumption for the SPECweb benchmark is more narrower than SPECpower 
as all load-levels of SPECweb consume power between 55 to 80 percent of the 
peak power limit possible. In general, there is less opportunity to limit the 
power consumption of memory than the package subsystem.

\emph{In summary, there is opportunity to limit the power consumption of SPECpower and SPECweb 
at different load-levels below the 50-ms resolution for the package and DRAM subsystems.}

\section{Efficacy of Power Limiting}
\label{sec:results}

In this section, we discuss the effects of power limiting on the 
performance and power of SPECpower and SPECweb benchmarks. Specifically, 
we investigate whether we can achieve energy-proportional operation for 
these benchmarks by leveraging the RAPL interfaces. Through our experiments, 
we show that most of the power savings comes from the PP0 domain and 
memory subsystem power limiting contributes the least to achieving power 
savings. 

\begin{figure*}[htb]
\centering
\includegraphics[scale=0.62]{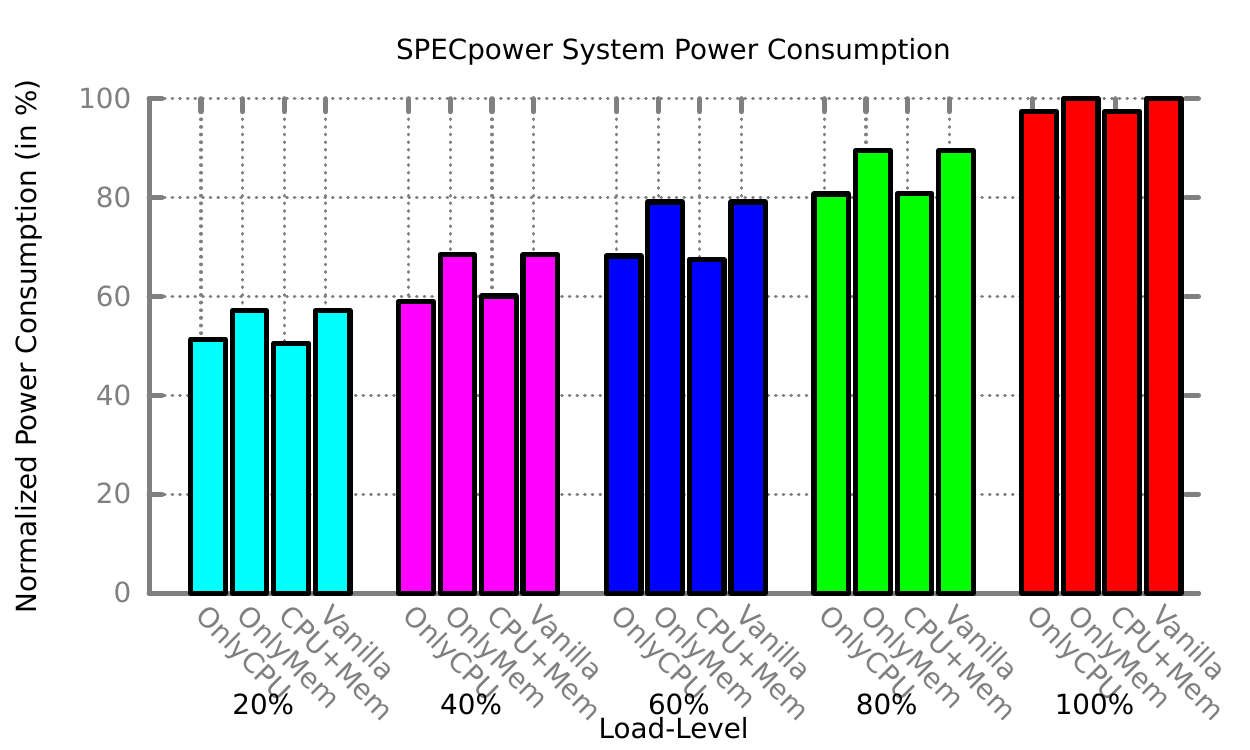}
\includegraphics[scale=0.62]{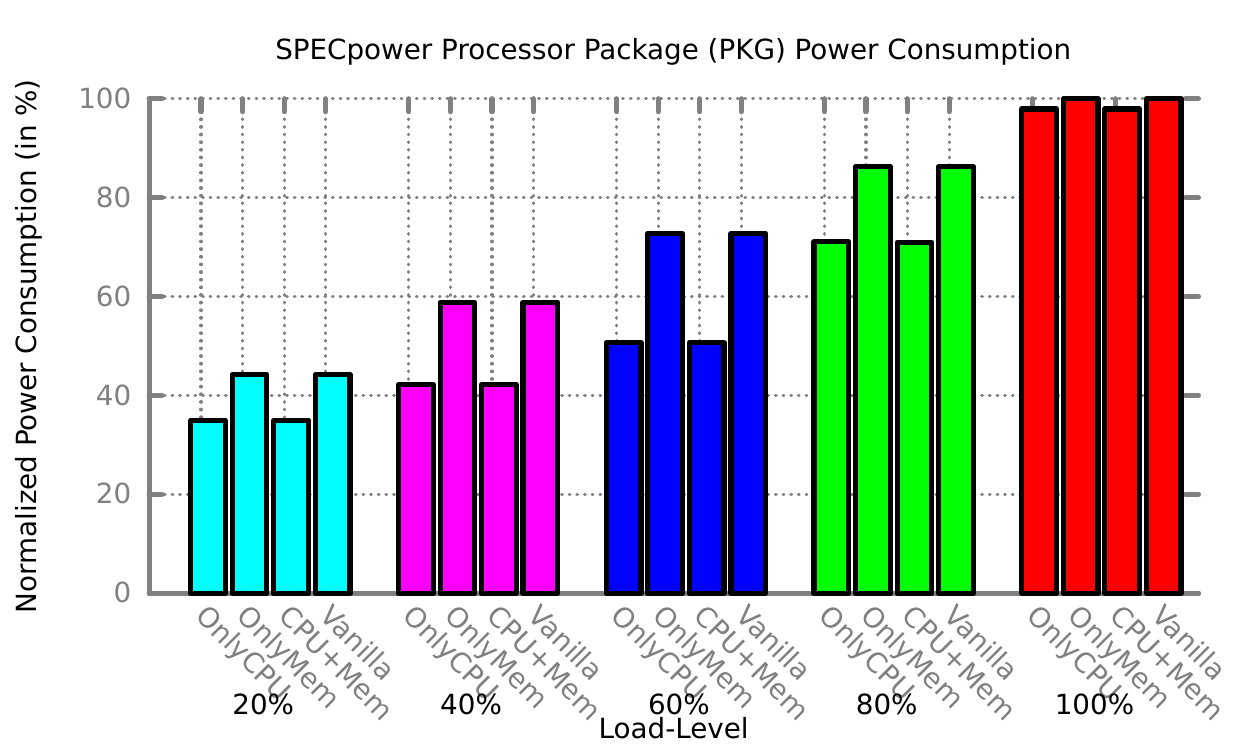}
\includegraphics[scale=0.62]{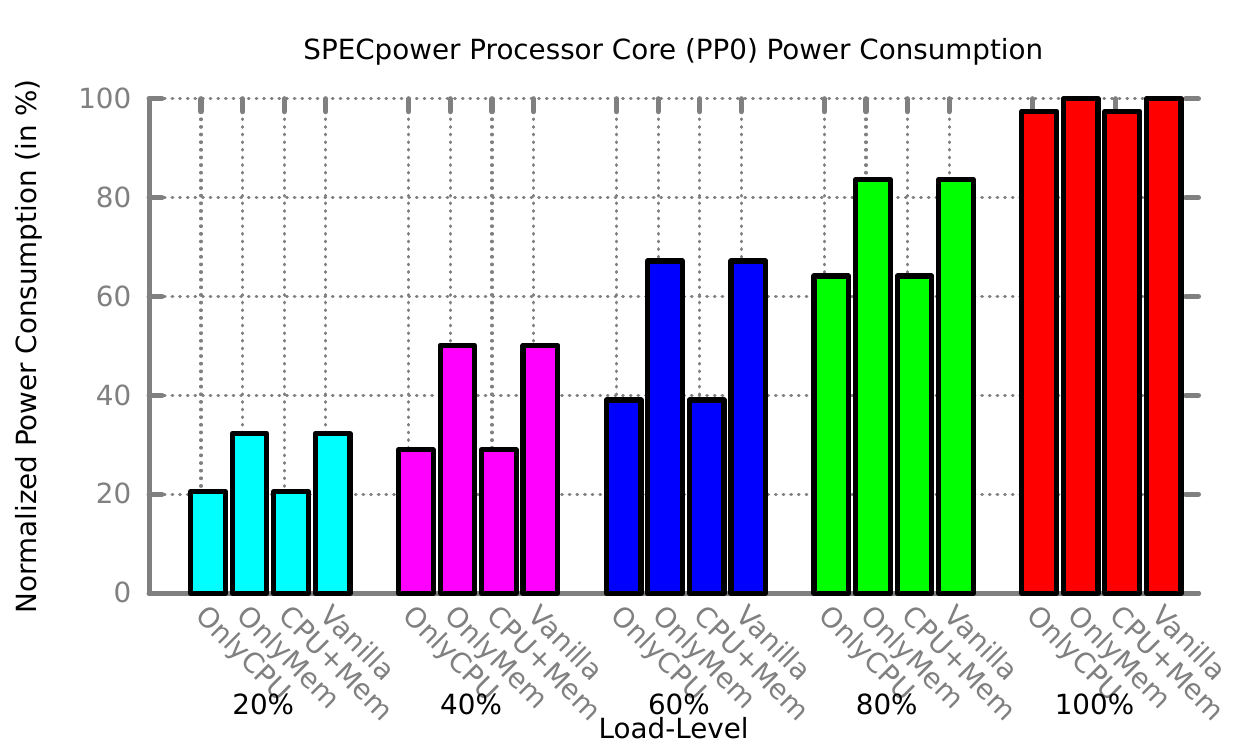}
\includegraphics[scale=0.62]{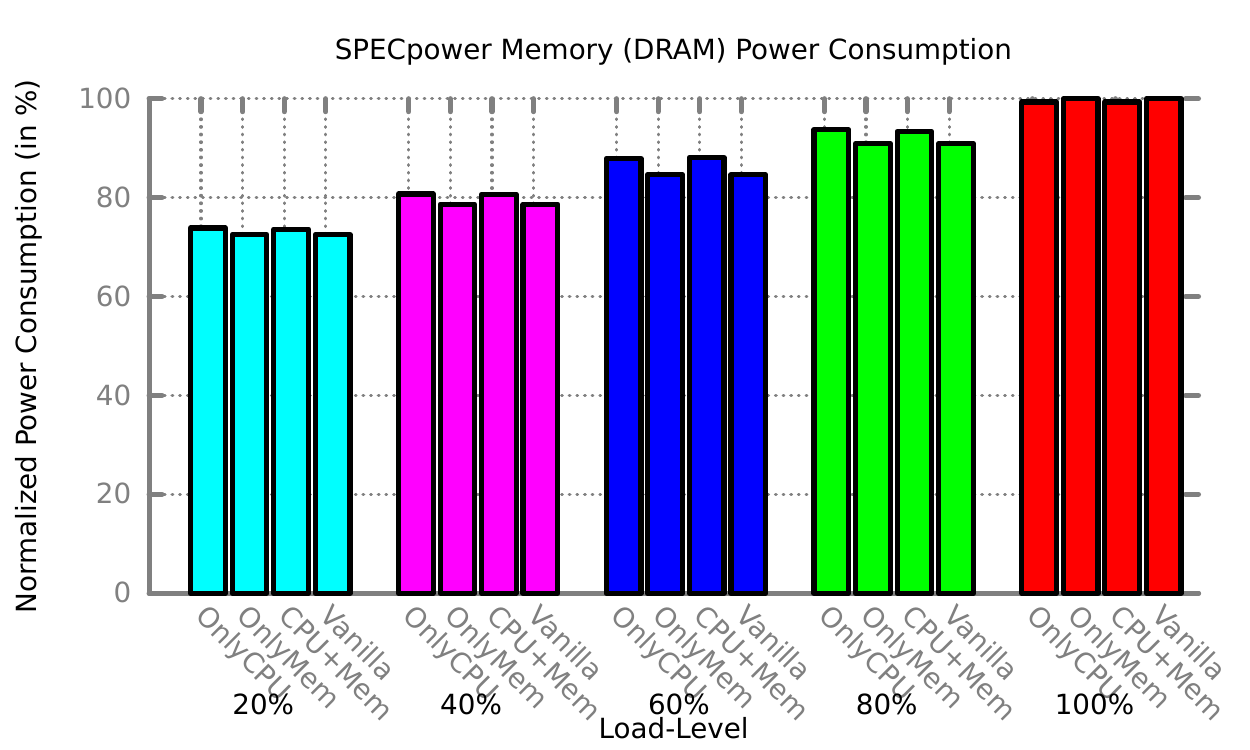}
\caption{Impact of Power Limiting on SPECpower}
\label{fig:pspecpower}
\end{figure*}

\begin{figure*}[htb]
\centering
\includegraphics[scale=0.62]{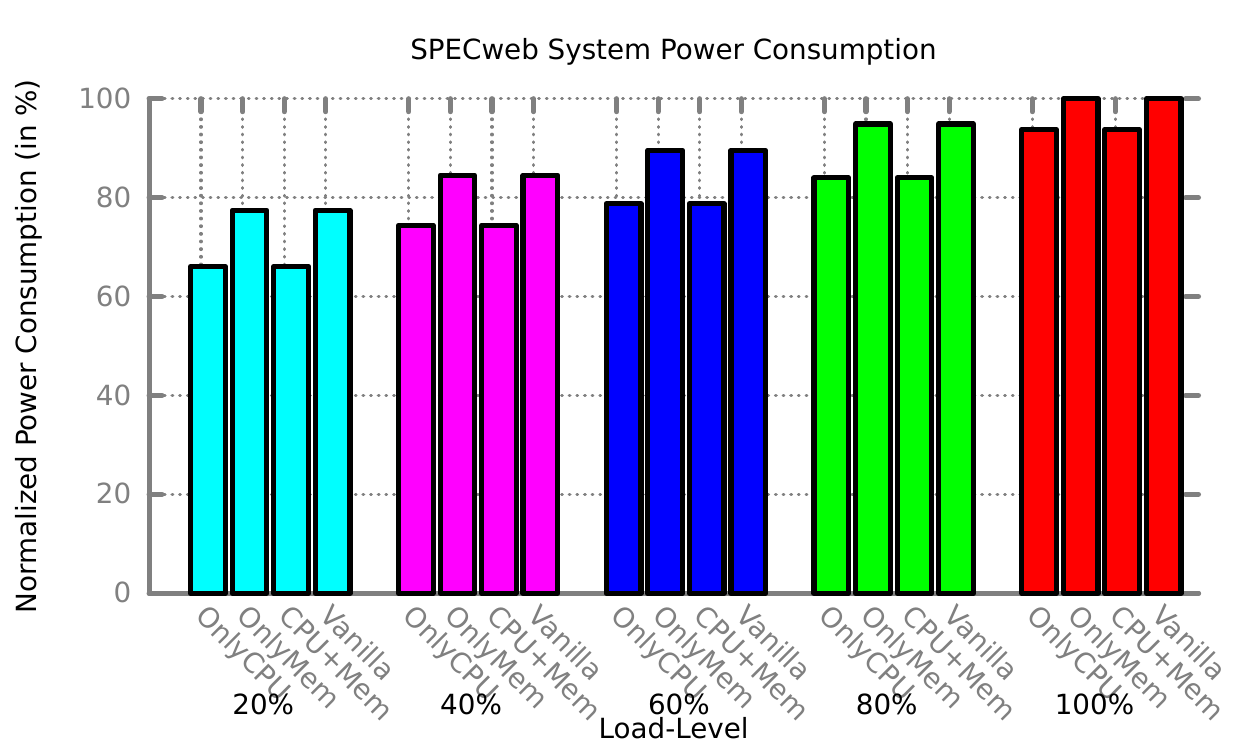}
\includegraphics[scale=0.62]{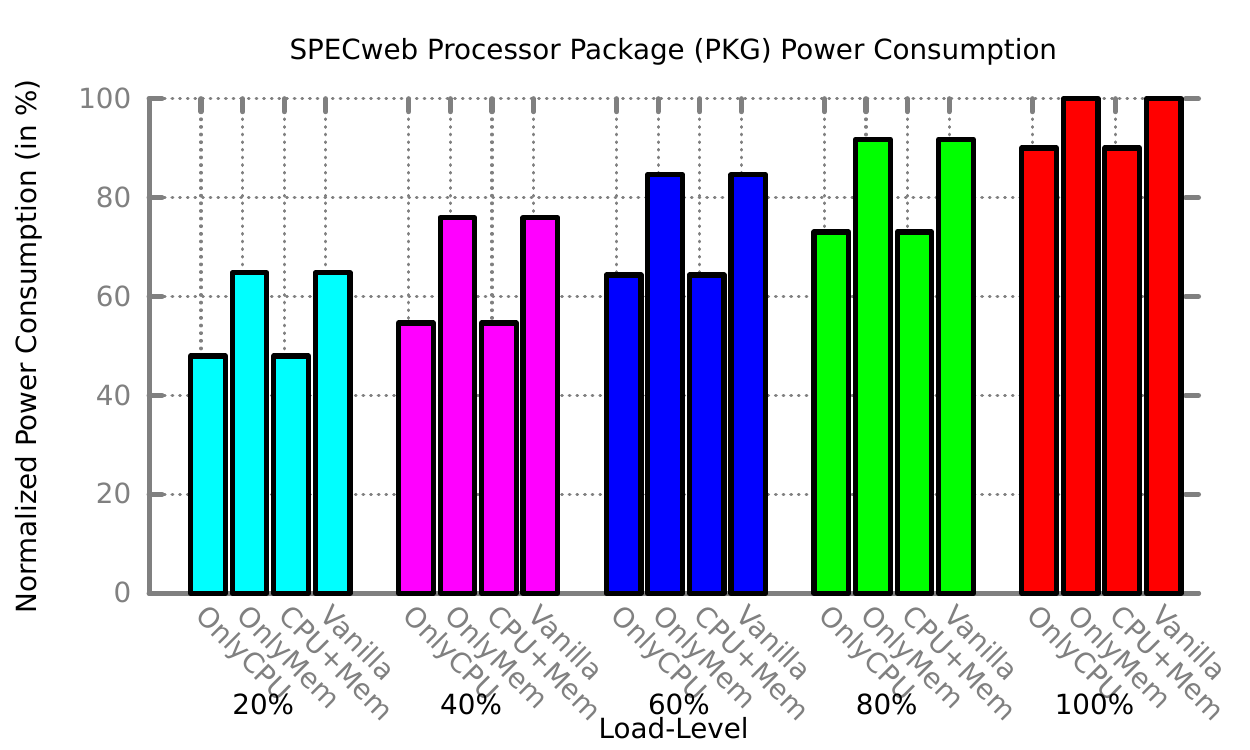}
\includegraphics[scale=0.62]{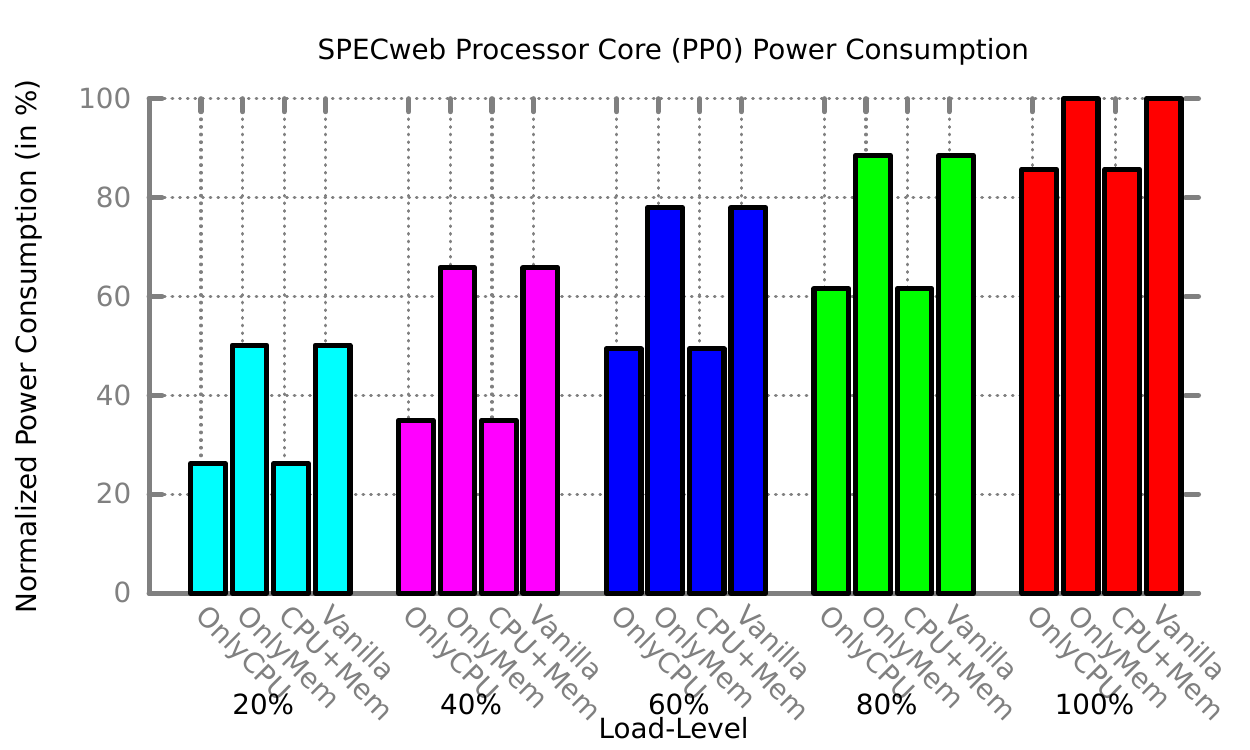}
\includegraphics[scale=0.62]{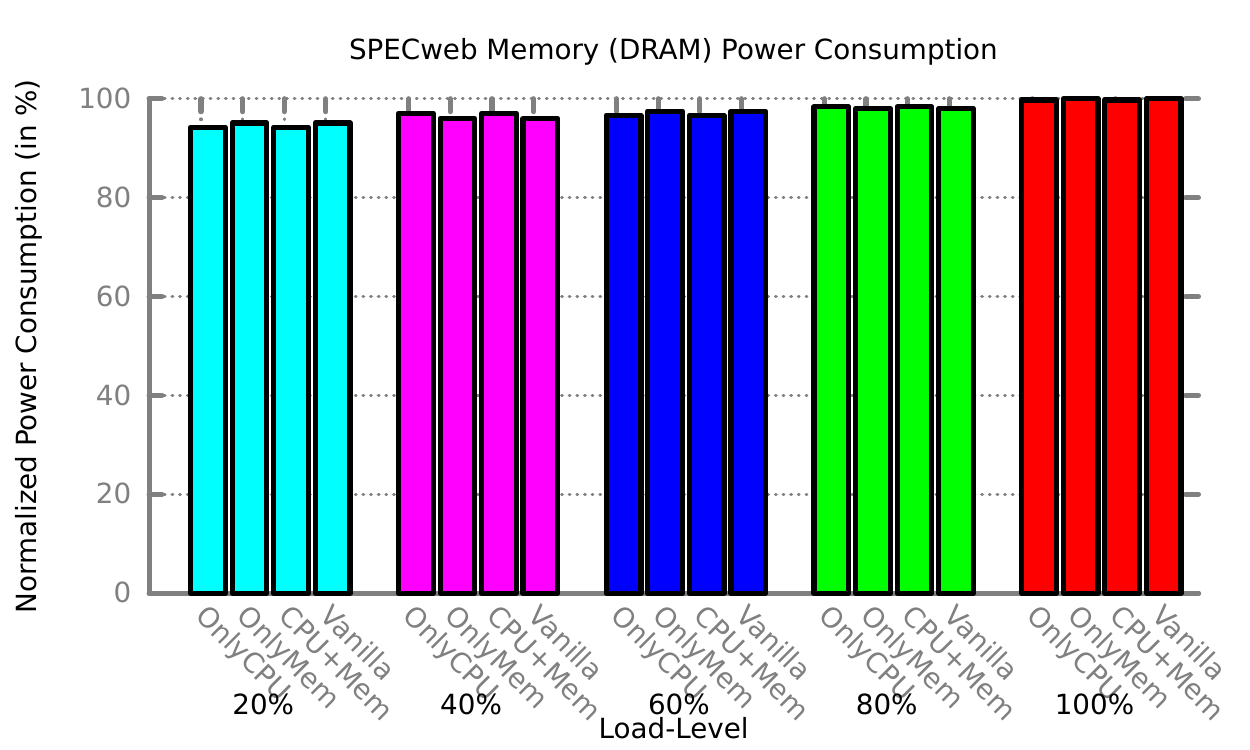}
\caption{Impact of Power Limiting on SPECweb}
\label{fig:pspecweb}
\end{figure*}

\begin{figure*}[htb]
\centering
\includegraphics[scale=0.6]{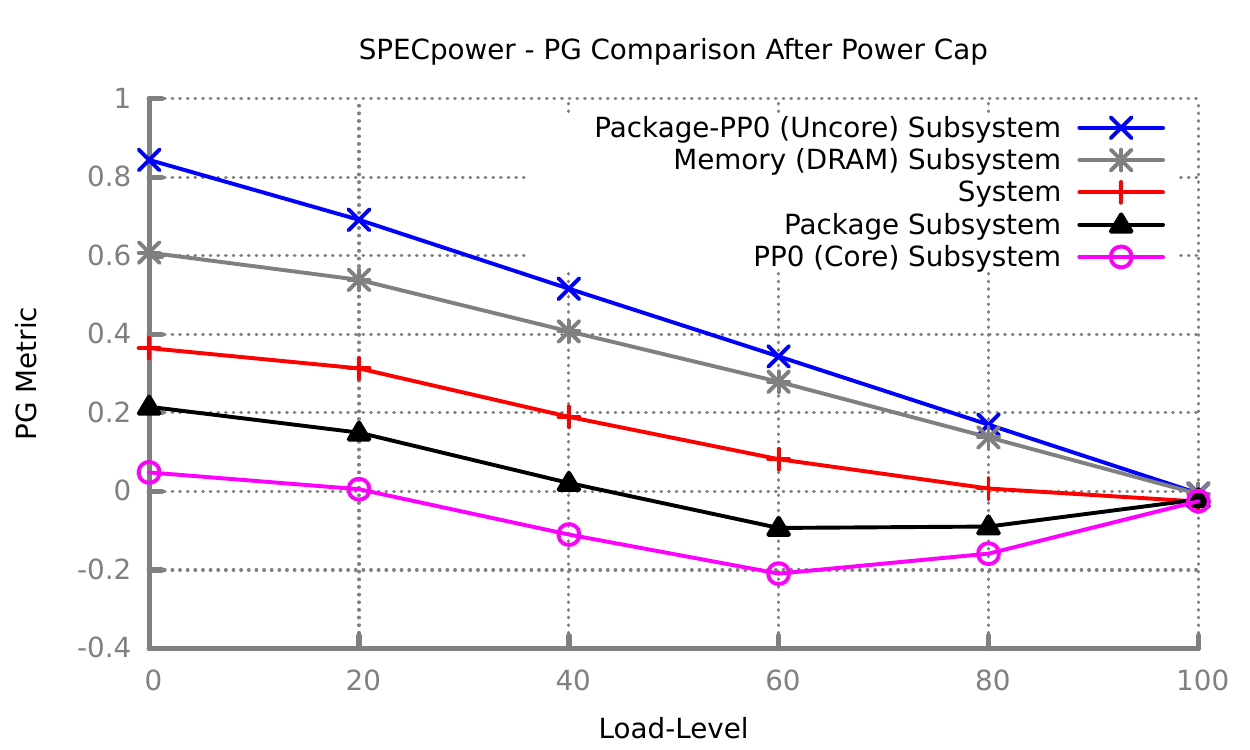}
\includegraphics[scale=0.6]{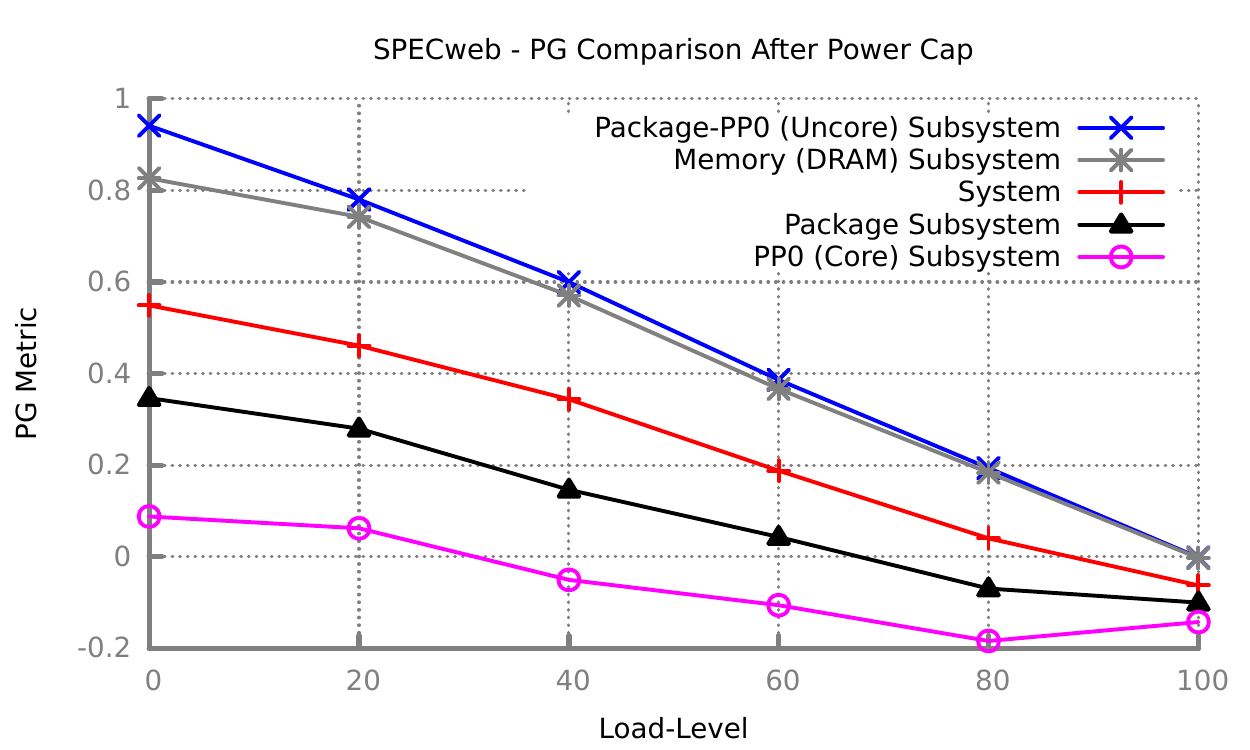}
\caption{PG Metric after Power Cap}
\label{fig:pgafteropt}
\end{figure*}

\begin{figure*}[htb]
\centering
\includegraphics[scale=0.6]{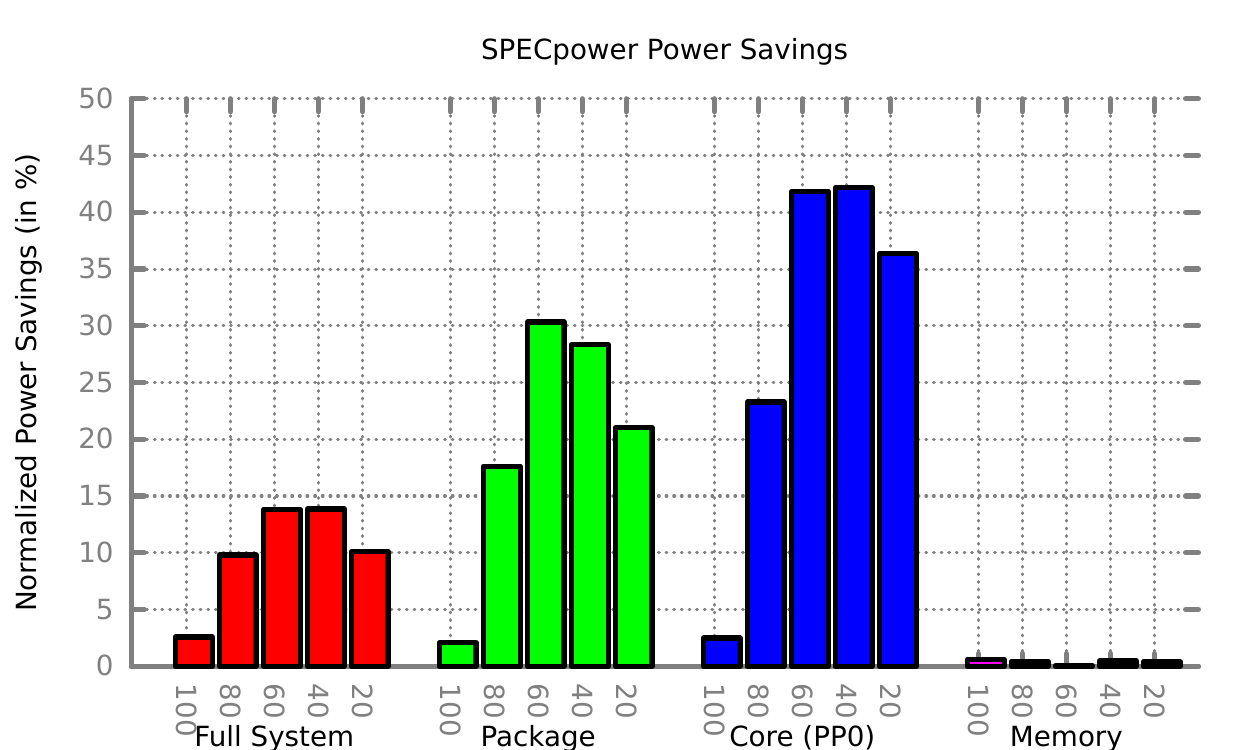}
\includegraphics[scale=0.6]{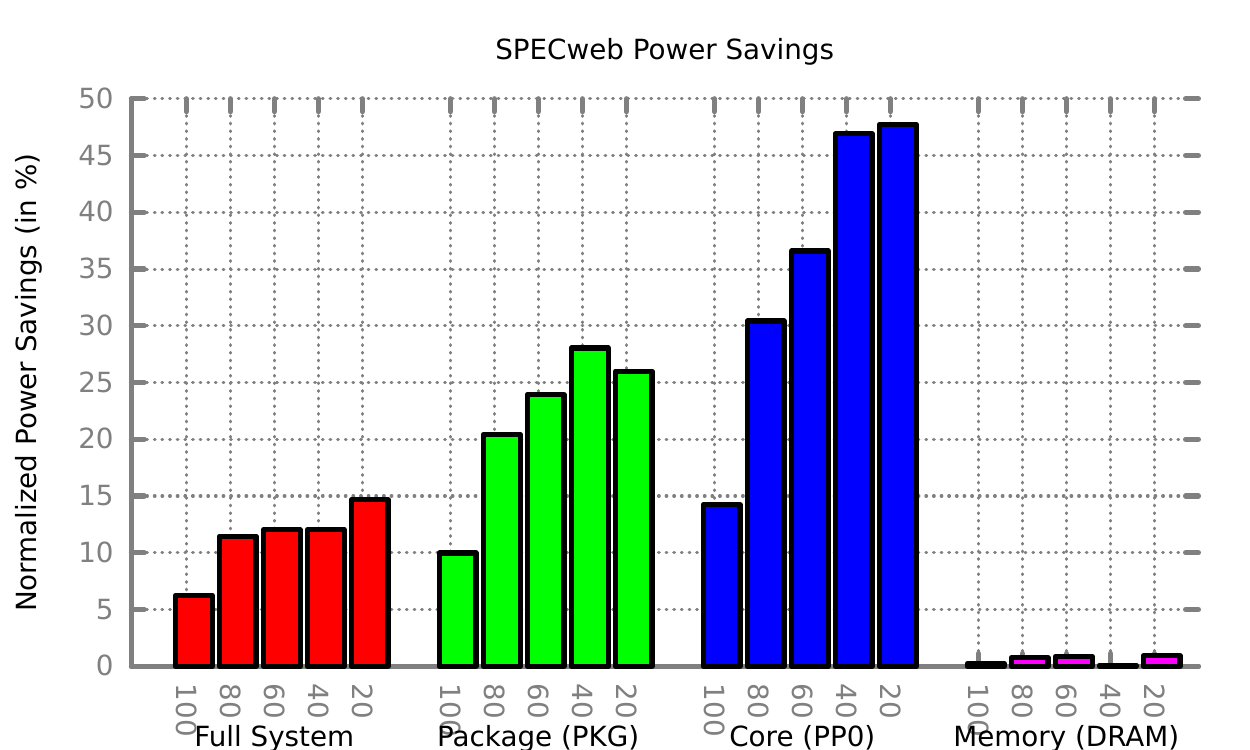}
\caption{Power Savings}
\label{fig:psavings}
\end{figure*}

\begin{figure*}[htb]
\centering
\includegraphics[scale=0.6]{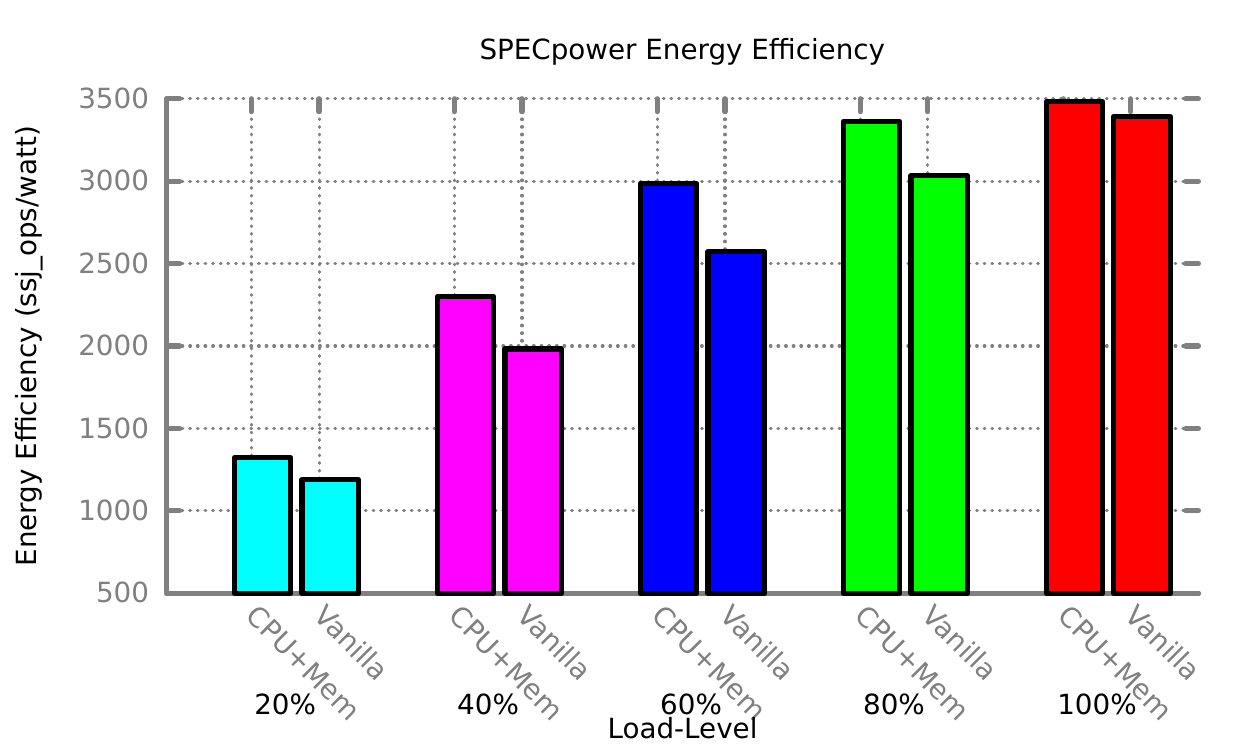}
\includegraphics[scale=0.6]{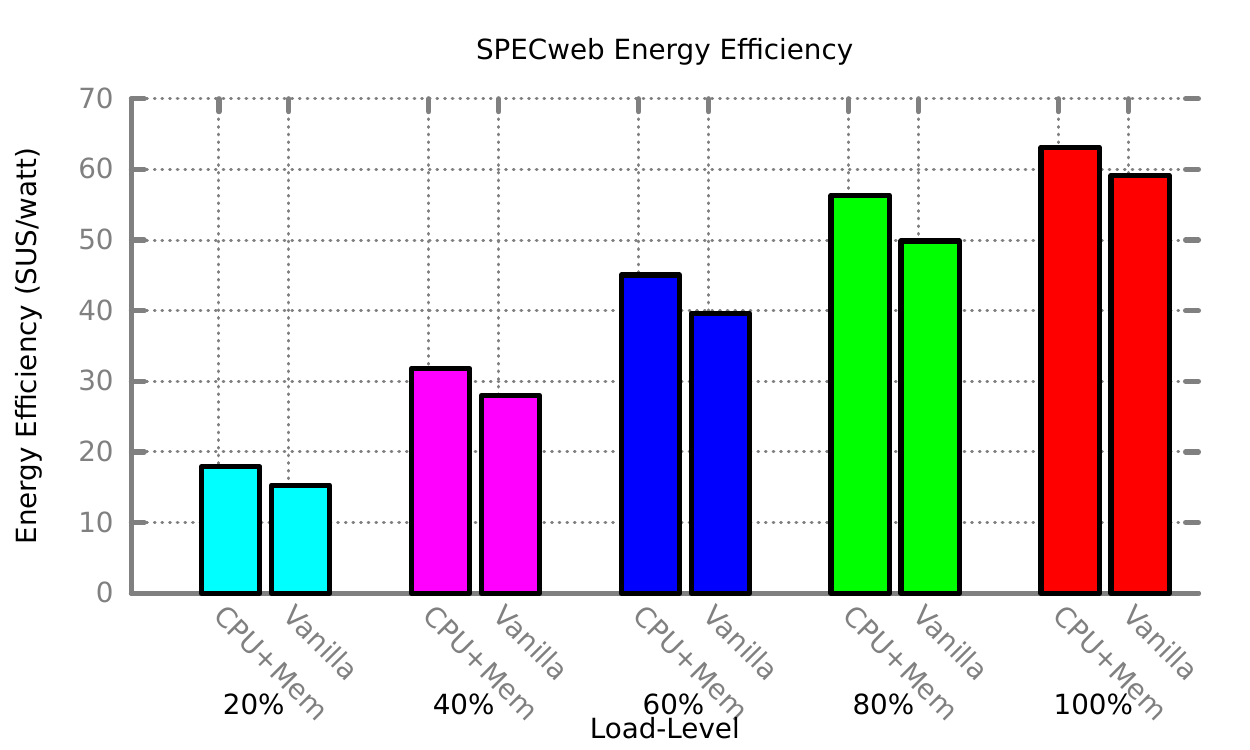}
\caption{Energy Efficiency}
\label{fig:eff}
\end{figure*}

\subsection{Methodology}

We run both the SPECpower and SPECweb benchmarks at five different 
load-levels (from 20\% to 100\% in steps of 20) under power limit. 
Our experiments focus on PP0 and DRAM power limiting. We don't focus on 
processor package power limiting as the \emph{uncore} subsystem does not contribute
to power savings at any load-level and all the power savings came from the 
PP0 domain while we experimented with processor package power limiting~\cite{icpe_eprop}. Our 
experiments present results for three different power limiting scenarios:

\begin{itemize}

\item \emph{CPUOnly} policy: Performance under only PP0 subsystem power limit.
\item \emph{MemOnly} policy: Performance under only DRAM subsystem power limit.
\item \emph{CPU+Mem} policy: Performance under PP0 and DRAM subsystem power limits.

\end{itemize}

In our experiments, we manually configure the power limit using RAPL interfaces.
For the CPUOnly and MemOnly policies, we manually set 15 different power 
limits below the average power consumption of the corresponding subsystem. 
These 15 different power limits start from the average power consumption to 28 
watts less than average power consumption at steps of 2 watt each. For the CPU+Mem 
policy we look at all possible power limits for a total of 225 combinations 
for each load-level. In this paper, we only present 
the best possible power savings \emph{without} performance degradation for the 
benchmarks. We only present runs which achieve performance within 1\% of target load-level for SPECpower. 
In case of SPECweb, we present results which achieve the target load-level and maintain 
\emph{TIME\_GOOD} and \emph{TIME\_TOLERABLE} constraints (see Section~\ref{sec:setup}). 
We also use the least possible value as the time window for power limiting 
(i.e., 976 microseconds). 

\subsection{Impact of Power Limiting}

Figure~\ref{fig:pspecpower} shows the normalized power consumption of five 
different load-levels of SPECpower. The values are normalized against the 
power consumption at 100\% load-level vanilla run. We show the power 
consumption for the full system (top left) and processor package (top right), 
PP0 (bottom left) and DRAM (bottom right) subsystems. Such 
representation of the power consumption allows to identify whether we 
achieve energy proportionality at a particular load-level. 

We observe that we achieve energy proportionality for the full system only for 
100\% and 80\% load-levels. However, power limiting reduces the power 
consumption for other load-levels even though we are not able 
to achieve energy-proportional operations. We would like to emphasize that 
the system consumes 36.51\% of peak power even when idling. The CPU+Mem and CPUOnly 
policies achieves the best power consumption. The MemOnly policy achieves 
negligible power consumption reduction. In case of processor package power consumption, 
we are able to achieve energy-proportional operation for all loads-levels 
except 20\% load-level. Moreover, the reduction in power consumption is more 
when compared to the full system. This is an expected outcome as we only have 
power limiting control over processor package, PP0 (which is a part of processor package) and 
DRAM subsystems. Through power limiting, we achieve energy-proportional operation 
for all load-levels when we look at PP0 domain in isolation. As mentioned earlier, 
we achieve negligible power reduction from the DRAM subsystems while meeting 
the performance constraints of the benchmarks. In case of the subsystems, the 
different power limiting policies have same effect as seen for the full 
system (i.e., using CPU+Mem and CPUonly results in best possible power 
reduction whereas using MemOnly results in negligible power savings). 

Figure~\ref{fig:pspecweb} shows the normalized power consumption of five 
different load-levels of SPECweb benchmark. Similar to SPECpower, we show the power 
consumption for the full system (top left) and processor package (top right), PP0 
(bottom left) and DRAM (bottom right) subsystems. We achieve 
energy-proportional operation only for 100\% load-level in case of SPECweb. 
Similar to SPECpower, we however achieve power savings for other 
load-levels. The power reduction for SPECweb is less than power reduction 
seen for SPECpower as SPECpower is \emph{more} energy-proportional than SPECweb 
(see Section~\ref{sec:avgpower}). We would also like to stress that the 
SPECweb benchmarks idles at 54.88\% of its peak power. When looking at the 
processor package power consumption in isolation, we are able to achieve 
energy-proportional operation for 100\% and 80\% load-levels. PP0 domain 
provides the highest power reduction and achieves energy-proportional 
operation for all load-levels except 20\% load-level. The memory subsystem 
does not contribute much to the power reduction. CPU+Mem and CPUOnly 
policies provide the best power reduction possible. 

Table~\ref{tab:epafteropt} shows the EP metric for the configuration 
which achieves best power savings 
of the SPECpower and SPECweb benchmarks. We see components with EP metric 
$>$ 1 indicating that we are operating at better than energy-proportional 
trade-off points. As expected, the EP metric for the 
PP0 subsystem has seen a substantial increase. For the PP0 domain, the metric 
increases from 0.85 to 1.18 and 0.63 to 1.12 for the SPECpower and SPECweb 
benchmarks respectively. The memory and the \emph{uncore} subsytem does see 
any significant EP metric improvement. 

\begin{table}[t]

\centering
\caption{Summary of Full System- and Subsystem-Level Energy Proportionality and Linear Deviation After Power Caps.}
\label{tab:epafteropt}

\begin{tabular}{|c|c|c|c|} 
\hline
\textbf{Subsystem}	&	\textbf{Benchmark} & \textbf{EP Metric}	& \textbf{LD Metric}	\\ \hline
\multirow{2}{*}{Full System}	&	SPECpower & 0.69	&	-0.044	\\ \hhline{~---}
								&	SPECweb & 0.48	& -0.024	\\ \hline
\multirow{2}{*}{Package (PKG)} 	&	SPECpower &	0.96	& -0.1490	\\ \hhline{~---}
								&	SPECweb & 0.79	& -0.101	\\ \hline
\multirow{2}{*}{\emph{Core} (PP0} 	&	SPECpower &	1.18	&	-0.221	\\ \hhline{~---} 
								&	SPECweb & 1.12	& -0..192	\\ \hline
\multirow{2}{*}{\emph{Uncore} (Package-PP0)} 	&	SPECpower &	0.16	& 0.004	\\ \hhline{~---}
												&	SPECweb & 0.03	& 0.019	\\ \hline
\multirow{2}{*}{Memory (DRAM)} 	&	SPECpower &	0.37	& 0.013	\\ \hhline{~---}
								&	SPECweb & 0.09	& 0.045	\\ \hline
\end{tabular}
\end{table}

Figure~\ref{fig:pgafteropt} shows the PG metric for the configuration which achieve 
best power savings. For the Package and PP0 subsystems, the PG metrics is negative 
for some load-levels suggesting that we achieve better than energy-proportional 
operation. As observed both memory and \emph{uncore} subsystem are not amenable 
to operating at different power performance trade-off points as the PG metric 
trend decrease linearly for those subsystems. 

Table~\ref{tab:epafteropt} also shows the LD metric for the best power 
savings run. We are able to shift the linear deviation from positive to 
negative for the full system, Package and PP0 subsystems. Our approach 
improves the energy proportionality of the server by improving the linear 
deviation of subsystems. 

\subsection{Power Savings}

Figure~\ref{fig:psavings} shows the power savings for the SPECpower (left) 
and SPECweb (right) benchmarks. Power limiting conserves between 3\% to 15\% 
of power at the full system-level. We would like to stress that the subsystems 
over which we don't have power limiting control consume between 11\% and 17\% power 
of the full system depending upon the load-level. The power savings at 100\% load-level for 
SPECpower is less than SPECweb as the former is a CPU-intensive benchmark 
and most of our power savings come from the PP0 domain. We observe that 
the memory subsystems provides negligible power savings. In case of PP0 
domain we conserve between 3\% and 30\% for SPECpower and 14\% to 45\% for 
SPECweb depending upon the load-level. 

\subsection{Impact on Energy Efficiency}

Figure~\ref{fig:eff} shows the energy efficiency of SPECpower and 
SPECweb at five different load-levels for the CPU+Mem and vanilla runs. The 
energy efficiency of SPECweb and SPECpower are represented as \emph{ssj\_ops/watt} 
and \emph{SUS/watt} respectively. The improvement is calculated as the ratio 
of difference between the energy efficiency under power limit and vanilla 
run over the vanilla run. We achieve energy efficiency improvements in 
all cases. SPECpower and SPECweb achieve up to 16 and 17 percent energy 
efficiency improvement, respectively, due to power limiting. 

\subsection{Impact on Instantaneous Power Consumption}

\begin{figure}[t]
\centering
\includegraphics[scale=0.6]{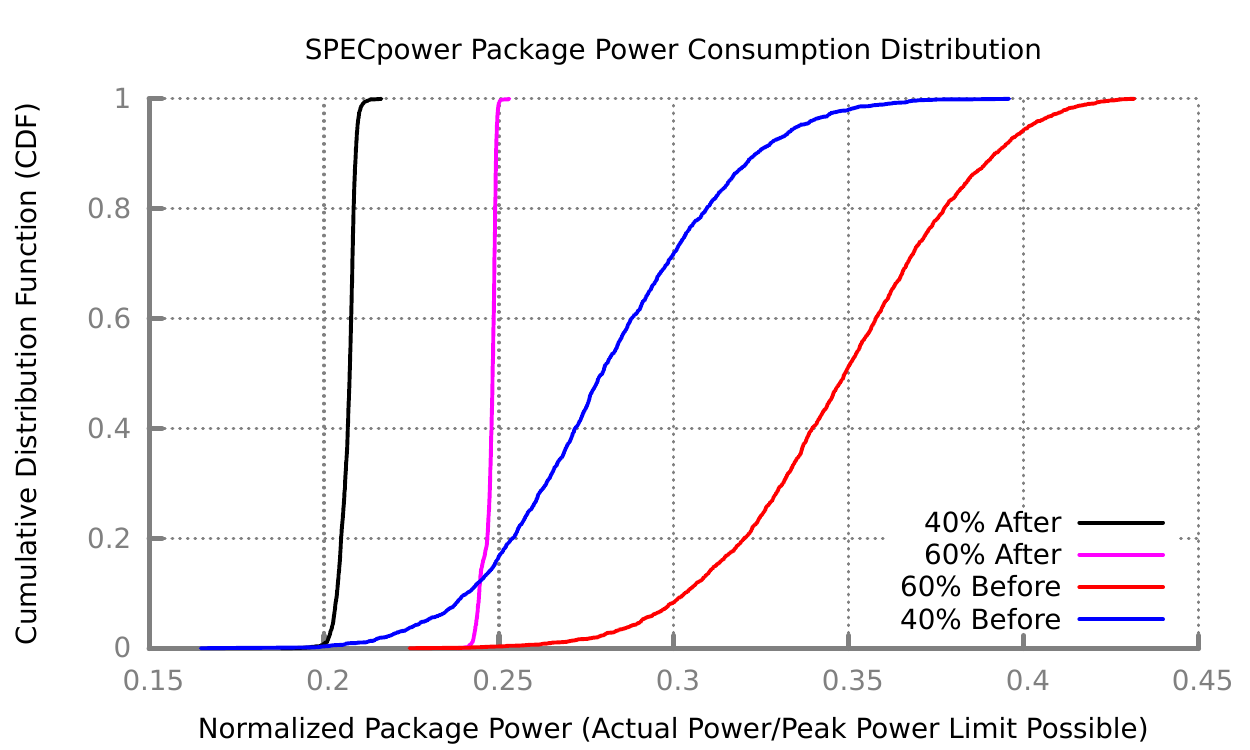}
\caption{Instantaneous SPECpower Package Power Consumption (After and Before Applying Power Caps)}
\label{fig:instoptpkg1}
\end{figure}

\begin{figure}[t]
\centering
\includegraphics[scale=0.6]{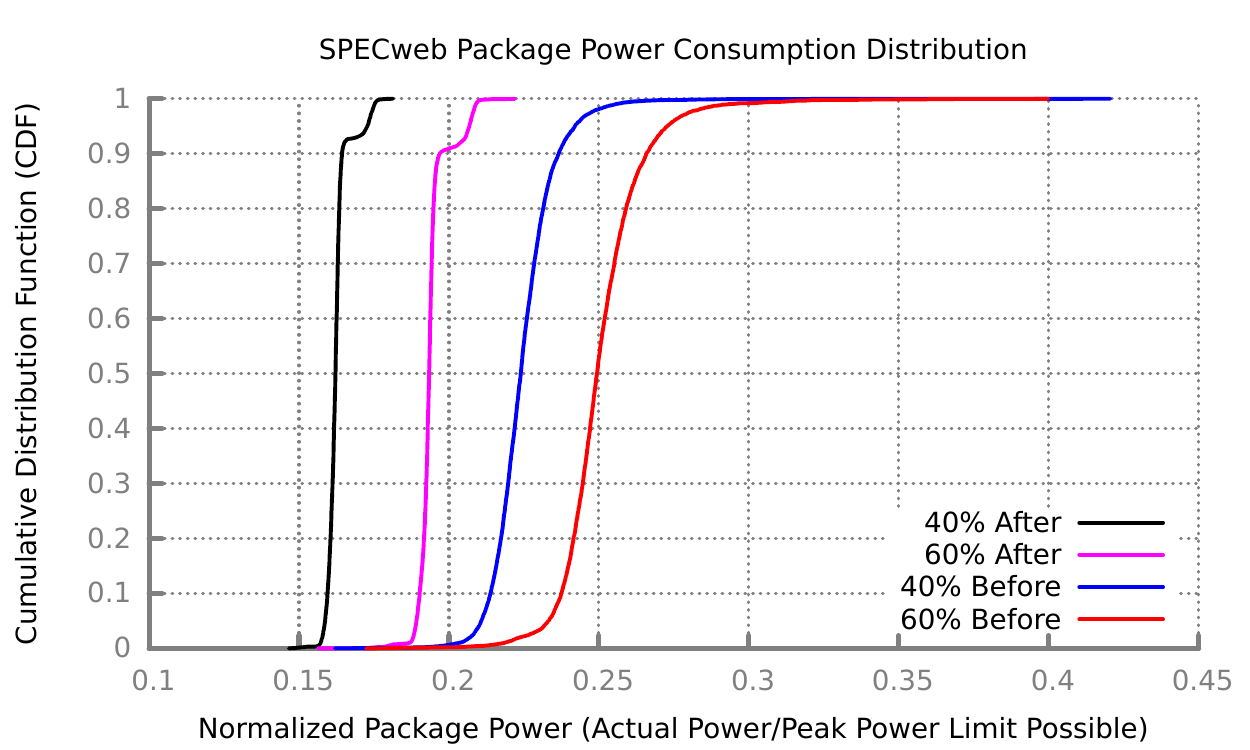}
\caption{Instantaneous SPECweb Package Power Consumption (After and Before Applying Power Caps)}
\label{fig:instoptpkg2}
\end{figure}

Over-provisioning leads to the wasting of infrastructure resources, and the 
maximum instantaneous power consumed by the subsystems is an important factor in 
determining the power budget for a system. Determining the optimal power 
provisioning strategy requires an understanding of the instantaneous 
power profile of the system. Towards this end, the instantaneous power 
profile is discussed in this section. We describe the difference in 
instantaneous power profile between two different load-levels 
(40 and 60 percent) with and without power cap for both SPECpower and 
SPECweb benchmarks. The instantaneous power profile for the 
configuration which achieved best power savings is shown. The power profile of the 
package and memory subsystems is collected at 50 ms resolution in all cases.

\begin{figure}[t]
\centering
\includegraphics[scale=0.6]{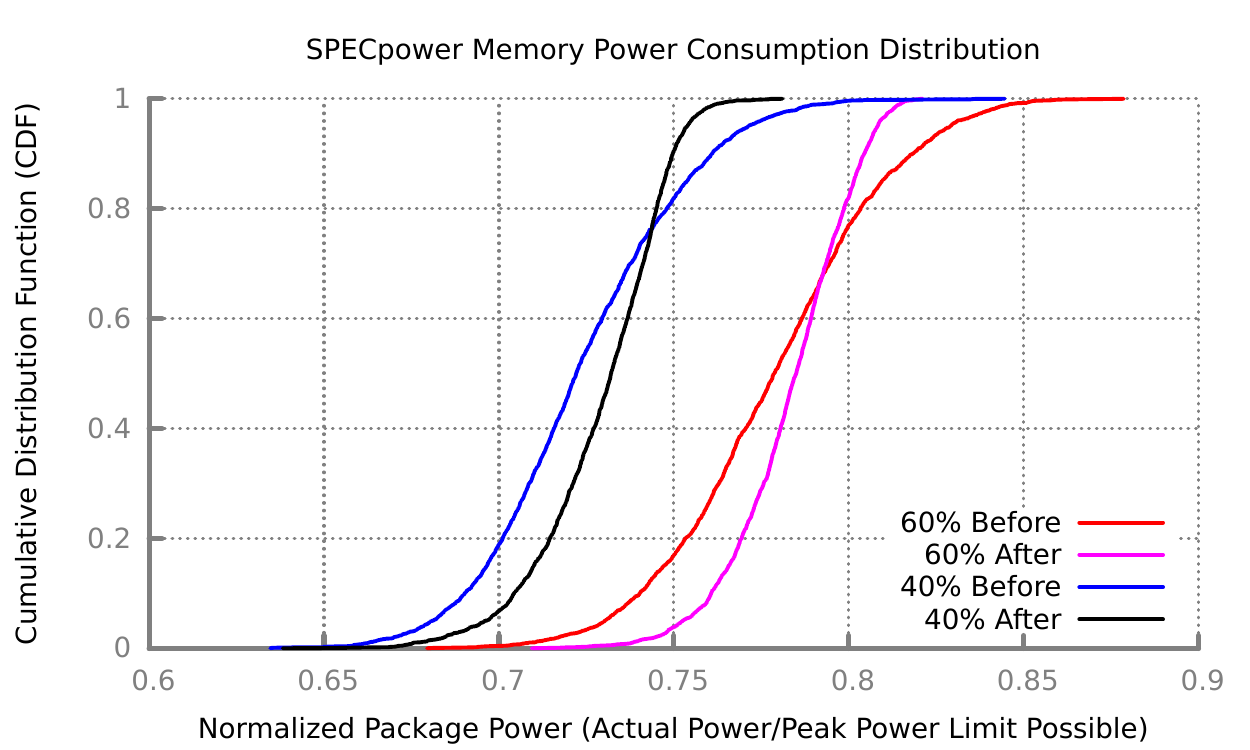}
\caption{Instantaneous SPECpower Memory Power Consumption (After and Before Applying Power Caps)}
\label{fig:instoptmem1}
\end{figure}

\begin{figure}[t]
\centering
\includegraphics[scale=.6]{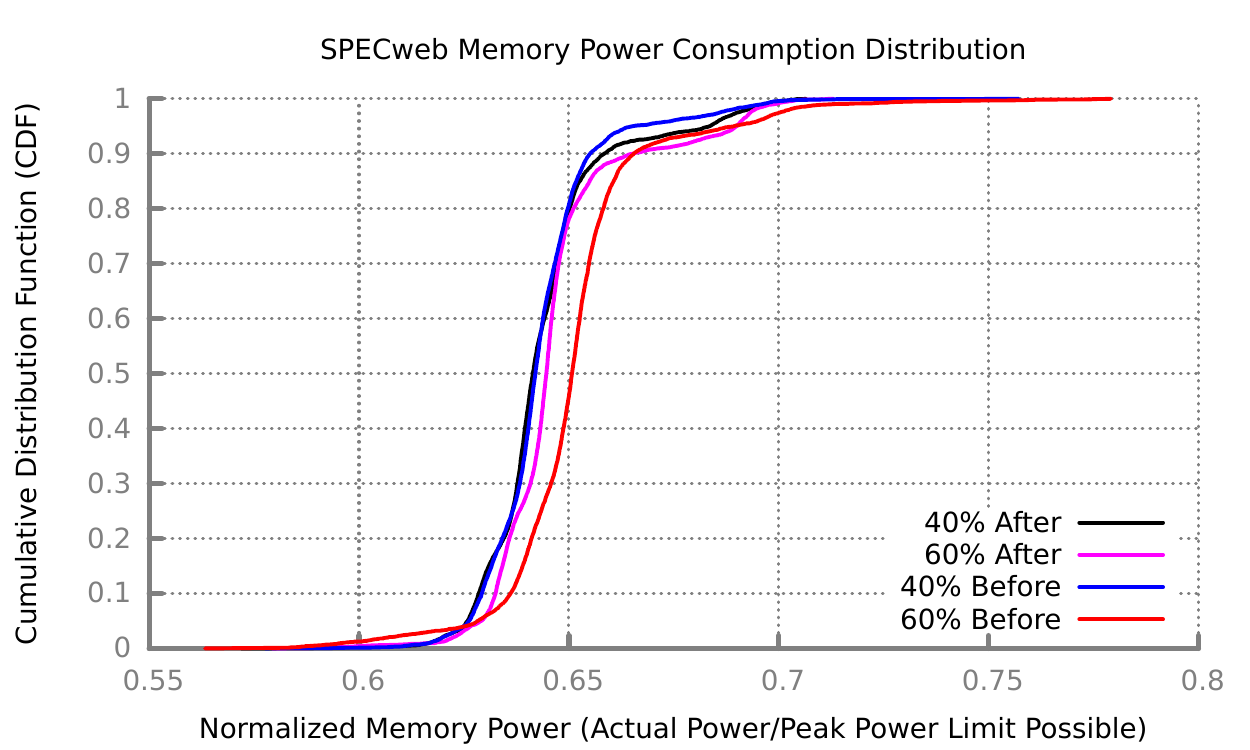}
\caption{Instantaneous SPECweb Memory Power Consumption (After and Before Applying Power Caps)}
\label{fig:instoptmem2}
\end{figure}

Figures\ref{fig:instoptpkg1} and~\ref{fig:instoptpkg2} show the instantaneous power profile of the 
package subsystem at 40 and 60 percent load-level with and without the 
power cap for SPECpower and SPECweb respectively. Power limiting works as expected for both SPECpower and 
SPECweb benchmarks. The range of instantaneous power consumption is narrowed 
due to power capping. Moreover, the power limiting removes the relatively 
few power spikes indicated by the flat lines at 100\% (see Section~\ref{sec:instpower}). 
Such power limiting mechanisms are useful for power provisioning without impacting the 
performance of the application.

Figures~\ref{fig:instoptmem1} and~\ref{fig:instoptmem2} show the instantaneous power profile of the 
memory subsystem at 40 and 60 percent load-level with and without the 
power cap for both the benchmarks. The relatively few power spikes in the memory subsystem for both 
the benchmarks are removed due to power limiting. Even though we don't achieve 
considerable power savings at the memory subsystem-level due to power limiting, 
applying appropriate power limits such that the impact on performance is controlled at 
desirable level can help make power provisioning decisions and increase the efficiency 
of the server.

\section{Related Work}
\label{sec:related}

\subsection{Energy-Proportional Operation For Enterprise Class Workloads}

Wong et al.~\cite{knightshift} provide an architecture for improving the energy 
proportionality using server-level heterogeneity. They combine a high-power compute node 
with a low-power processor essentially creating two different power-performance operation 
regions. They save power by redirecting requests to the low-power processor at 
low request rates thereby improving energy proportionality. Our work looks at improving the 
energy proportionality of traditional servers by improving the subsystem-level 
energy proportionality using RAPL interfaces.  

Meisner et al.~\cite{oldi} characterize online data-intensive services 
(OLDI) to identify opportunities for power management, design a framework that predicts
the performance of OLDI workloads and investigate the power and
performance trade-offs using their simulation framework. Fan et al.~\cite{provisioning} 
investigate the benefits of
energy-proportional systems in improving the efficiency of power
provisioning using their power models. They provide evidence
that energy-proportional systems will enable improved
power-capping at the data-center level. In contrast, we look at
leveraging the \emph{power-capping} mechanism to achieve
energy-proportional operation for SPECpower and SPECweb. 

Tolia et al.~\cite{ensemble} proposed that by migrating workloads from
under-utilized systems to other systems and turning the under-utilized
systems off, energy proportionality can be approximated at an
ensemble-level (i.e., for a group of nodes or rack-level). They used
virtual machine (VM) migration as a software mechanism to move
workloads off of under-utilized systems. In this paper, we use
user-defined and hardware-enforced power limiting to achieve
energy-proportional systems at the node-level.

\subsection{Subsystem-level Power Management}

Deng et al.~\cite{coscale} propose the CoScale framework which dynamically adapts the frequency 
of the CPU and memory respecting a certain application performance degradation target. 
They also take per-core frequency settings into account ~\cite{multiscale}. Li et al.~\cite{crosscomponent} study the 
CPU microarchitectural adaptation and memory low power states to reduce energy 
consumption of applications bounding the performance loss by using a slack allocation algorithm. Our paper deals 
with subsystem-level power management on a \emph{real system}. 

\subsection{Power Limiting}
Several mechanisms to \emph{cap} the power consumption of the system have 
been studied~\cite{packcap,capidle}. However, we study the use of RAPL power limiting 
which is hardware-enforced in this paper. David et al.~\cite{rapl} 
proposed RAPL and evaluated RAPL for the memory sub-system. 
They present a model that accurately predicts the power consumed by the DIMMs 
and use RAPL to \emph{cap} the power consumption. Rountree et al.~\cite{barrypowerlimit} 
use RAPL power limiting to study the behavior of performance for benchmarks 
in the NAS parallel benchmark suite. Specifically, they are interested in the 
performance of various compute nodes under a power bound. Weaver et al.~\cite{papipower} have 
have exposed RAPL energy meters through PAPI. We use RAPL interfaces to achieve energy-proportional 
operation for SPECpower and SPECweb benchmarks and to the best of our knowledge, there is no 
previous study on using RAPL interfaces for enterprise class server workloads.

\section{Conclusion}
\label{sec:conclusion}
The management of power and energy is a key issue for data centers. Efficient 
power management of enterprise-class server workloads have the potential to 
greatly reduce energy-related costs and facilitate efficient power 
provisioning. 

Energy proportionality holds the potential to significantly 
improve the energy efficiency of data centers. Consequently, in this paper, 
we investigate the potential of achieving energy proportionality for
SPECpower and SPECweb benchmarks using RAPL interfaces. Our study sheds light on
the mechanisms for power management of enterprise-class server
workloads and the efficacy of RAPL interfaces. We identify the least
and most energy-proportional subsystem using the on-chip energy
meters. We then characterize the instantaneous power profile of these 
benchmarks to identify if there is any opportunity to limit the power 
consumption of these benchmarks. Finally, we present our results on the 
impact of power limiting on the power, performance and energy efficiency of 
SPECpower and SPECweb benchmarks. Our results show that we are able to 
achieve power savings of up to 15\%.

\appendices

\section{SPECpower Batch Size has No Effect on Power Consumption}
\label{sec:app1}

\begin{figure}[htb]
\centering
\includegraphics[scale=.63]{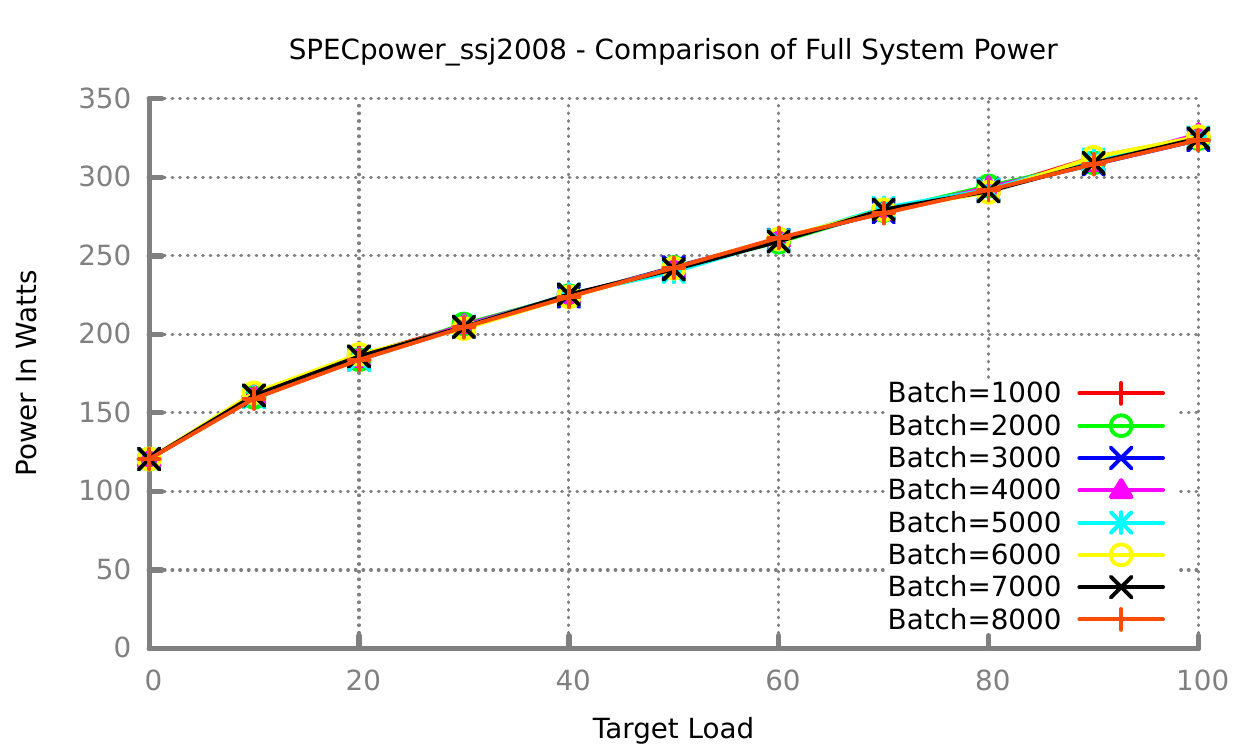}
\caption{Average Power Consumption of SPECpower}
\label{fig:avgpower}
\end{figure}

Figure~\ref{fig:avgpower} shows the average power consumption of SPECpower 
benchmark at different load-levels. For full-system power measurement, we
have followed the power measurement methodology specified and
developed by the SPEC organization for the SPECpower
benchmark~\cite{SPECpowermethod}. The figure also shows the effect of 
changing the batch sizes in SPECpower. We wanted to quantify this effect as 
batching queries to exploit and create opportunities for power management 
is a well-researched area~\cite{powernap}. The number of
transactions in each batch scheduled in SPECpower benchmark is
calibrated using the \textit{input.scheduler.batch\_size} input
parameter. We use eight different batching sizes from 1000 to 8000 in steps of
1000. Each data presented is the average of 10 runs. The standard deviation for 
the power consumed during the runs were less than $\pm$2\% of the mean.

Our results shed light on the repeatability of our experiments and the 
consistency of SPECpower benchmark. We observe that the lines in the plot 
overlap each other. Based on our experiments, the batch sizes have minimal 
to no effect on the power consumed by the benchmark. Similarly, changing 
batch sizes did not have any effect on the power consumption of subsystems 
(i.e., package, core and memory) as well. 

\section{SPECweb at 13000 SUS is Network-Intensive}
\label{sec:app2}

\begin{figure}[htb]
\centering
\includegraphics[scale=.63]{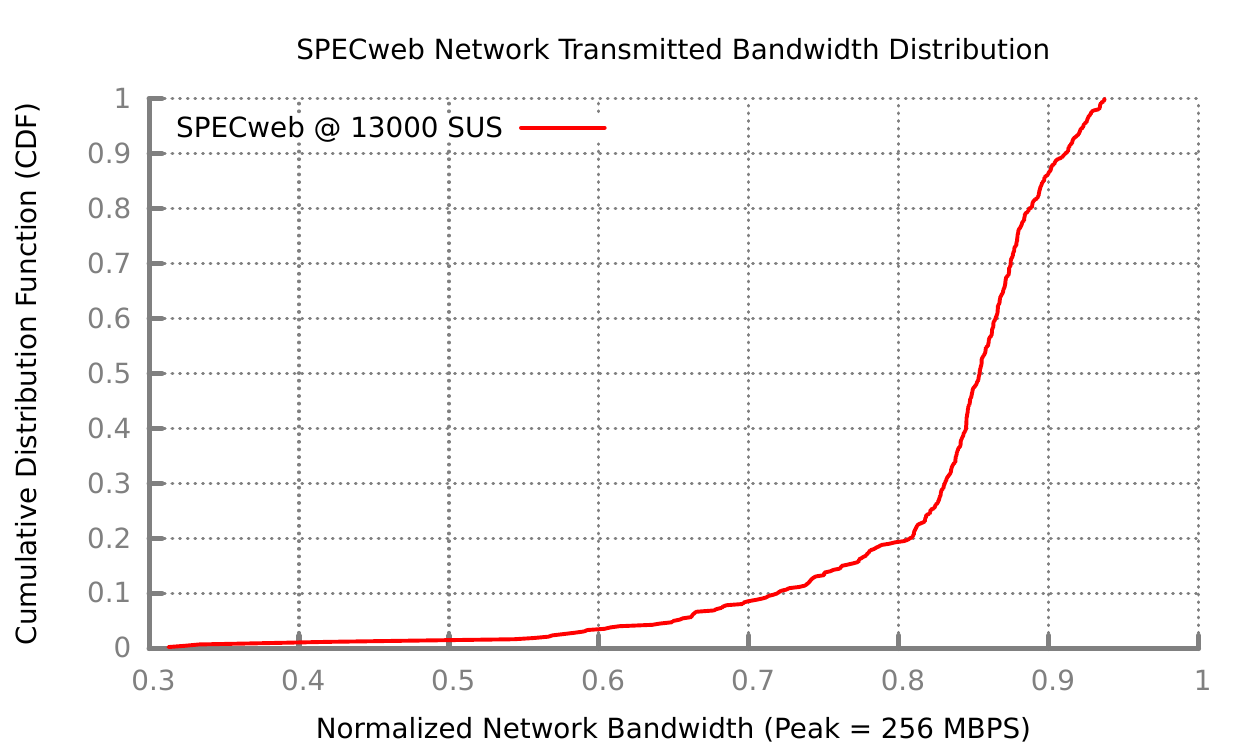}
\caption{CDF of SPECweb Network Bandwidth}
\label{fig:netbw}
\end{figure}

Figure~\ref{fig:netbw} shows the cumulative distribution function (CDF) for 
the transmitted network bandwidth while running SPECweb at 13000 SUS. 
This CDF presents the percentage of total time where the transmitted bandwidth was either at or below a certain 
percentage of peak bandwidth possible. In our case the peak bandwidth is 
256 megabytes per seconds (MBPS) due to the bonded Ethernet connection on the 
testbed (see Section~\ref{sec:setup}). We monitor the network bandwidth 
using the \emph{sar} utility at a resolution of one second. We observe 
that the SPECweb benchmark at 13000 SUS is networking intensive. The benchmark 
spends 80\% of the time consuming more than 80\% of the 
network bandwidth. Moreover, it spends 50\% of time 
consuming more than 85\% of the network bandwidth. Through our experiments 
we also found that the system under test was not able to meet the response 
time constraints when we increased the SUS beyond 13000. Hence, our 
experiments use 13000 SUS as 100\% load-level for SPECweb.

\bibliographystyle{abbrv}	
\bibliography{rapl_ccpe_journal}
\vspace{-15mm}
\begin{IEEEbiographynophoto}{Balaji Subramaniam}
is a Ph.D. student in Computer Science department at 
Virginia Tech. His research interests 
include energy-proportional computing, power modeling and prediction, 
hardware- and software-controlled power management, and benchmarking.
\end{IEEEbiographynophoto}

\vspace{-15mm}
\begin{IEEEbiographynophoto}{Wu-chun Feng} is Professor and 
Elizabeth and James E. Turner Fellow of Computer Science at Virginia Tech. 
He received his Ph.D. in Computer Science from the University of Illinois 
at Urbana-Champaign in 1996.
\end{IEEEbiographynophoto}

\end{document}